\RequirePackage{fix-cm}
\documentclass[smallextended]{svjour3}       
\smartqed  
\usepackage[T1]{fontenc}
\usepackage[utf8]{inputenc}
\usepackage[english]{babel}
\usepackage{amssymb}
\usepackage{amsmath}
\usepackage{latexsym}
\usepackage{mathtools}
\usepackage{graphicx}
\usepackage{bm}
\usepackage{amsbsy}
\usepackage{wasysym}
\usepackage{booktabs}
\usepackage{caption}
\usepackage{tabularx}
\usepackage{epsfig}
\usepackage{hyperref}

\DeclarePairedDelimiter{\abs}{\lvert}{\rvert}

 \journalname{Celestial Mechanics and Dynamical Astronomy}
\begin{document}

\title{A semi-analytical model of the Galilean satellites' dynamics
}
%



\author{Giacomo Lari   
}


\institute{Giacomo Lari \at
              Dipartimento di Matematica, Università di Pisa,\\
              Via Bruno Pontecorvo 5, Pisa, Italy \\
              Tel.: +39-0502213263\\
              Fax: +39-0502210678\\
              \email{lari@mail.dm.unipi.it}           
}

\date{Received: February 21, 2018 / Accepted: June 25, 2018}

\maketitle

\begin{abstract}
The Galilean satellites' dynamics has been studied extensively during the last century. In the past it was common to use analytical expansions in order to get simple models to integrate, but with the new generation computers it became prevalent the numerical integration of very sophisticated and almost complete equations of motion. In this article we aim to describe the resonant and secular motion of the Galilean satellites through a Hamiltonian, depending on the slow angles only, obtained with an analytical expansion of the perturbing functions and an averaging operation. In order to have a model as near as possible to the actual dynamics, we added perturbations and we considered terms that in similar studies of the past were neglected, such as the terms involving the inclinations and the Sun's perturbation. Moreover, we added the tidal dissipation into the equations, in order to investigate how well the model captures the evolution of the system.
\keywords{Laplace resonance \and tidal dissipation \and secular model}
\end{abstract}

\section{Introduction}
Since 1610, year of the discovery made by Galileo Galilei, the Galilean satellites have fascinated many astronomers and scientists. They were the first clear example of objects orbiting around a body, Jupiter, different from the Sun or the Earth. Moreover, they form a miniature copy of the Solar System, which could be observed easily with the first telescopes. Already in 1798, Laplace showed in~\cite{laplace} that some of these satellites were resonant. In fact, the motion of Io, Europa and Ganymede, is characterized by a three-body mean motion resonance with ratio $4:2:1$, today known as Laplace resonance. It means that the period of Ganymede around Jupiter is (almost) two times the one of Europa and (almost) four times the one of Io. Callisto, the fourth and furthest from Jupiter, is not resonant, although its mean motion is near the $3:7$ commensurability with Ganymede.

\begin{table}[t]
{\small
\begin{center}
\caption{Overview of some Galilean satellites' physical and mean orbital parameters at J2000. The first are taken from~\cite{schub}, while the second are obtained with a digital filtering of the JUP310 ephemerides (\cite{jup310}).}\label{tab:galsat}
\noindent\begin{tabular}{lcccc}
\toprule
 & Io & Europa & Ganymede & Callisto\\
\midrule
\textbf{Physical parameters} & & & & \\
Mass ($10^{22}$ $kg$)		& $8,932$ & $4,800$ & $14,819$ & $10,759$ \\
Mean radius ($10^3$ $km$)	& $1,821$ & $1,565$ & $2,631$ & $2,410$\\
\textbf{Orbital parameters} & & & & \\
Period ($days$)  & $1,7691$ & $3,5512$ & $7,1546$ & $16,6890$ \\
Mean motion ($^\circ/days$)  & $203,49$ & $101,37$ & $50,32$ & $21,57$\\
Semi-major axis ($10^5$ $km$)  & $4,220$ & $6,713$ & $10,706$ & $18,831$\\
Eccentricity  & $0,0042$ & $0.0095$ & $0.0013$ & $0,0074$\\
Inclination ($^\circ$)  & $0,04$ & $0,46$ & $0,21$ & $0,20$\\
\bottomrule
  \end{tabular}
\end{center} }
\end{table}

Mean motion resonances are not uncommon for asteroids and satellites, but most of the times they involve two bodies only. The Galilean satellites were the first known case in the Solar System of a three-body resonance, although recently it was discovered a similar resonance between the satellites of Pluto, as reported in~\cite{pluto}.

The Laplace resonance is a very strong configuration: Io, Europa and Ganymede are deeply locked in resonance, in particular there are some combinations of the orbital angles (mean longitudes and longitudes of the pericenters) that librate around fixed values, with very small oscillations. This means that some orbital configurations occur often (with respect to their orbital motion), modifying significantly the orbits.

One of the most important effects of the resonance is that it forces the eccentricities to higher values, with respect to their free values, which are almost $0$. The combined effect of the resonances with the tides acting on the satellites produces a significant dissipation inside the moons, especially within Io that is closer to the planet. Studying the amount of the dissipated energy, in~\cite{peale} it was foreseen the presence of volcanoes on Io and in~\cite{cassen} it was proposed the existence of an ocean of liquid water under the icy crust of Europa.

Moreover, the Laplace resonance allows to spread the huge dissipation within Io to all the three satellites involved in the resonance. In fact, the dissipative effects acting on Io do not change its orbit only, but also the ones of Europa and Ganymede. This mechanism makes the resonance evolve, but it is still not sure toward which future.

In literature it is quite common the use of analytical and semi-analytical models for the study of the Galilean satellites' resonant dynamics. In~\cite{desitter} the author investigated the possible existence of periodic orbits in the system: this special case is called de Sitter resonance, which is different from the current Laplace resonance. In~\cite{malho} and~\cite{yope} the authors developed semi-analytical models for the study of the system's evolution due to the tidal dissipation, proposing possible origins of the Laplace resonance. More recently, in~\cite{paita}, the authors studied the configurations near the de Sitter resonance and their relation with the current resonance of the system.

However, none of the mentioned works performed an accurate comparison with numerical models or ephemerides. One of the reasons can be the difficulty to replicate correctly all the details of the dynamics, in particular the frequencies and the amplitudes of the resonant angles' libration. Differences in these quantities can lead to large discordances with the ephemerides.

In this article we present a new semi-analytical model, which contains the main effects acting on the system and describes the resonant and secular evolution of all the orbital elements of the four moons. With this model we obtain a good representation of the actual motion of the satellites, as showed by comparisons with numerical models. Moreover, we are able to reproduce quite well the migration of the orbits due to the dissipative effects.

The paper is structured as follows: in Section 2 we recall the Hamiltonian formalism; in Section 3 we introduce the perturbations we add to the model; in Section 4 we describe the operation of averaging that removes the short period terms from the Hamiltonian; in Section 5 we pass to slow variables and in Section 6 we present the results we obtain by a numerical integration of the model and we compare them with the filtered series of the moons' ephemerides. In Section 7 we add the dissipative effects due to the tides between Io and Jupiter and we quantify the variation in the Galilean satellites' semi-major axes. Finally, in Section 8 we summarize the results we achieved and we present the advantages of the model and possible applications.

\section{Hamiltonian theory}
Since we use the Hamiltonian formalism, it is useful to recall its main properties. In general, we consider a function $\mathcal H(\mathbf{p},\mathbf{q})$, called Hamiltonian, that depends on momenta $\mathbf p$ and coordinates $\mathbf{q}$. From this function, we can define a system of first order differential equations
\begin{equation}
\label{hamequ}
\begin{cases}
\mathbf{\dot p}=\displaystyle-\frac{\partial \mathcal H}{\partial \mathbf{q}}, \\
\mathbf{\dot q}=\displaystyle\frac{\partial \mathcal H}{\partial \mathbf{p}},
\end{cases}
\end{equation}
called Hamilton's equations. From~\eqref{hamequ} it is easy to see that the Hamiltonian is a first integral of the motion.

We recall also the notion of canonical transformations. They are special variables transformations
\[\phi:(\mathbf{p},\mathbf{q})\mapsto(\mathbf{P},\mathbf{Q}),\]
such that the Hamilton's equations relative to the new Hamiltonian
\[\mathcal H'(\mathbf{P},\mathbf{Q})=\mathcal H(\mathbf p(\mathbf{P},\mathbf{Q}),\mathbf q(\mathbf{P},\mathbf{Q}))\]
express the same dynamics of the initial differential equations.

In the case of a system of $N+1$ bodies with mass $m_i$ and barycentric positions $\mathbf z_i$ ($i=0,N$), the Poincaré variables are a convenient set of canonical coordinates. They consist in relative positions $\mathbf r_i=\mathbf z_i-\mathbf z_0$ and barycentric momenta $\mathbf p_i=m_i\mathbf {\dot z}_i$. As shown in~\cite{fmello}, the Hamiltonian can be reduced to
\begin{equation}
\label{poinham}
\mathcal H=\sum_{i=1}^N\Bigl(\frac{\abs{\mathbf p_i}^2}{2\beta_i}-\frac{\mu_i\beta_i}{\abs{\mathbf{r}_i}}\Bigr)+\sum_{i\ne k}\Bigl(-\frac{Gm_im_k}{\abs{\mathbf{r}_{ik}}}+\frac{\mathbf p_i\cdot \mathbf p_k}{m_0}\Bigr),
\end{equation}
where $G$ is the gravitational constant, $\mu_i$ is the gravitational parameter $G(m_0+m_i)$, $\beta_i$ is the reduced mass $m_0m_i/(m_0+m_i)$, and $\mathbf r_{ik}=\mathbf r_k-\mathbf r_i$.

In order to develop a secular theory, we want to pass from Cartesian coordinates to orbital elements. They are the semi-major axis $a$, the eccentricity $e$, the inclination $I$, the longitude of the pericenter $\varpi$, the longitude of the node $\Omega$ and the mean longitude $\lambda$. Since they are not canonical variables, we prefer Delaunay variables $(L,G,H,-\varpi,-\Omega,\lambda)$. The momenta are defined as:
\begin{align*}
&L_i=\beta_i\sqrt{\mu_ia_i},\\
&G_i=L_i(1-\sqrt{1-e_i^2}),\\
&H_i=G_i(1-\cos(I_i)).
\end{align*} 

In the case of the Galilean satellites $N=4$ and the central body is Jupiter. The unperturbed Hamiltonian, which describes the two-body motion between the satellites and Jupiter, is the first sum of~\eqref{poinham} and in the new variables becomes
\[\mathcal H_0=-\sum_{i=1}^4\frac{\mu_i^2\beta_i^3}{2L_i^2}.\]
It is clear that for $\mathcal H=\mathcal H_0$, only the variables $\lambda_i$ change, with constant rates called mean motions $n_i=\mu_i^2\beta_i^3/L_i^3$. This is far from the actual motion of the Galilean satellites.

In the next sections we will consider an equatorial reference system; this means that the plane $xy$ corresponds to the equatorial plane of Jupiter.

\section{Perturbations included in the model}
In the works we mentioned in the introduction, the authors developed a dynamical model for the three inner Galilean satellites and they assumed a planar motion. Since we want a model that approximates as well as possible all the orbital elements, we include Callisto in the model and we consider also nodes and inclinations of the moons. Moreover, we add terms which were not considered in the cited articles, like the gravitational perturbation of the Sun.

For a detailed analysis of the most important forces acting on the Galilean moons, we refer to~\cite{laineymodel}. In order of importance we have:
\begin{itemize}
\item the perturbation due to the coefficient $J_2$ of the Jovian gravitational field;
\item the mutual perturbations between the Galilean satellites;
\item the third-body perturbation of the Sun;
\item the perturbation due to the coefficients $J_4$ of the Jovian gravitational field.
\end{itemize}
All the other effects are far smaller than the ones mentioned, therefore a numerical model with just these effects is a good approximation of complete ephemerides.

First we introduce the effect of the oblateness of Jupiter. The gravitational potential can be expanded in series of Legendre polynomials $P_l$:
\begin{equation}
\label{harm}
U_J(\mathbf r_i)=-\frac{Gm_0}{\abs{\mathbf{r}_i}}\sum_{l=0}^\infty J_l\Bigl(\frac{R_0}{\abs{\mathbf{r}_i}}\Bigr)^lP_l(\sin\phi_i),
\end{equation}
where $R_0$ is the equatorial radius of Jupiter (almost $71398$ $km$) and $\phi_i$ is the moon's latitude (in the equatorial system). In our model we consider only the zonal harmonics $J_l$ with $l=2$ and $l=4$, whose values are $1.4736\times 10^{-2}$ and $-5.87\times 10^{-4}$ respectively (numbers taken from~\cite{gravjup}).

Then we consider the mutual perturbations between the Galilean moons: they are the most interesting forces, as they contain the resonant effects. In the Cartesian reference system they are described by the third body accelerations 
\[\mathbf {\ddot{r}}_i=Gm_k\Bigl(\frac{\mathbf r_{ik}}{\abs{\mathbf{r}_{ik}}^3}-\frac{\mathbf r_k}{\abs{\mathbf{r}_k}^3}\Bigr),\qquad i=1,4,\quad k\ne i.\]
The first addendum in the bracket is the direct part, while the second one is the indirect part. The associated potential is
\begin{equation*}
U_{3b}(\mathbf r_i)=Gm_k\Bigl(\frac{1}{\abs{\mathbf{r}_{ik}}}-\frac{\mathbf r_i\cdot\mathbf r_k}{\abs{\mathbf{r}_k}^3}\Bigr).
\end{equation*}
It is worth noting that in~\eqref{poinham} the indirect term comes out from the kinetic part of the Hamiltonian.

The perturbing function between the bodies $i$ (inner) and $k$ (outer) can be expanded in a Fourier series (as described in~\cite{fmello} or~\cite{murder}, Chapter 6)
\begin{equation}
\label{3body}
R^{(i,k)}=-\frac{Gm_im_k}{a_k}\sum_{\mathbf j}C_{\mathbf j}(a_i,a_k,e_i,e_k,I_i,I_k)\cos(\mathbf j\cdot \boldsymbol \Theta),
\end{equation}
where $\mathbf j=(j_l)_{l=1,6}$ is an integer vector and $\boldsymbol\Theta=(\lambda_i,\lambda_k,\varpi_i,\varpi_k,\Omega_i,\Omega_k)$, so that inside the brackets of the cosine there is a generic linear combination of the orbital angles of the two bodies. Not all the arguments are allowed in the expansion, since they must verify the D'Alambert rules. Moreover,~\eqref{3body} is an expansion in eccentricities and inclinations, that in the case of the Galilean satellites are quite small (see Table~\ref{tab:galsat}). Therefore, we do not need to truncate the series to a high order to obtain a good approximation.

The quantities $C_{\mathbf j}$ have as main term
\[c_{\mathbf j}(a_i/a_k)e_i^{\abs{j_3}}e_k^{\abs{j_4}}s_i^{\abs{j_5}}s_k^{\abs{j_6}},\]
where $s_i=\sin(I_i^2/2)$ (analogous relation for $s_k$) and the functions $c_{\mathbf j}$ are combinations of Laplace coefficients, which can be easily computed by an integral formula or a sum of a series. The expressions for all the coefficients are taken from~\cite{murder}, Appendix B.

The perturbations included so far are enough for studying the resonant motion of the moons, but if we want to compare the model with a complete numerical integration we lack one significant acceleration. In fact, for the outer satellites, the contribution of the Sun is essential to have a good representation of the motion. In particular, the inclinations of Ganymede and Callisto suffer greatly the Sun's perturbation and if we do not include it, we will get results completely different from the actual dynamics. In order to add this perturbation, we assume a simple two-body dynamics between the Sun and Jupiter. As our origin is fixed in the Jupiter's center of mass, we see the Sun orbiting around the planet. Then we can treat it as another satellite, very distant from its primary and that is not perturbed by the Galilean moons. The shape of the perturbative function due to the Sun is the same as~\eqref{3body}; however, we will have to pay a particular attention when we will pass to the resonant dynamics, because of the different time scales of the orbits.

In the end we have a Hamiltonian composed by the main part $\mathcal H_0$ and the perturbation $\epsilon \mathcal H_1$, where $\epsilon$ is the small parameter, that in the case of the mutual perturbations is the ratio $m_i/m_0$.

\section{Averaging and resonant model}
The perturbations presented in the previous section are the instantaneous forces that act on the Galilean satellites. They contain short period effects of the order of the orbital periods of the moons, i.e. few days. These effects make the orbital elements change wildly and fast, but they do not contribute in the long period dynamics. We want to remove all the short period terms from the Hamiltonian, in order to obtain a dynamics that depends only on the slow angles. This is possible through a process called averaging, which is equivalent to apply a mean over the fast angles to the Hamiltonian.

We use an approach based on Lie series. They allow to eliminate the short period terms from the expansion, through a process that is a canonical transformation (see~\cite{milknez90}). We define a function $\chi=\epsilon\chi_1+\epsilon^2\chi_2+...$ that determines a canonical transformation such that the new Hamiltonian is $\mathcal H'=\mathcal H'_0+\epsilon \mathcal H'_1+\epsilon^2 \mathcal H'_2+\epsilon^3 \mathcal H'_3+...$, where (using Poisson brackets)
\begin{align}
\label{newh}
\mathcal H'_0= & \mathcal H_0, \notag \\
\mathcal H'_1= & \mathcal H_1-\{ \mathcal H_0, \chi_1 \}, \notag \\
\mathcal H'_2= & -\{\mathcal H_0, \chi_2\}-\{\mathcal H_1, \chi_1\}+1/2\{\{\mathcal H_0, \chi_1\},\chi_1\}, \notag \\
\mathcal H'_3= & \dots
\end{align}
In this way all the short period terms are removed from $\mathcal H_1$, obtaining a new perturbative function $\mathcal H'_1$ with only secular terms. For example, if we want to remove the generic term $C_{\mathbf j}\cos(\mathbf j\cdot \boldsymbol \Theta)$, we have to add to $\chi_1$ the term $C_{\mathbf j}\sin(\mathbf j\cdot \boldsymbol \Theta)/(j_1n_1+j_2n_2)$.

As shown in~\eqref{newh}, this operation generates a new term $\mathcal H'_2$; since it is of the second order in the small parameter $\epsilon$, we can neglect it in our approximated model.

Apart from modifying the Hamiltonian, we are also changing the variables. In particular, when we conclude our operation of averaging, removing all the fast angles, we have new variables called mean orbital elements.

The purpose of this procedure is to eliminate all the terms containing the mean longitudes, as their variation is faster than that of the other angles ($\varpi_i$ and $\Omega_i$, which vary only under the action of the perturbations). However, because of the resonances, we have that the combinations $2\lambda_2-\lambda_1$, $2\lambda_3-\lambda_2$ and $4\lambda_3-\lambda_1$ have smaller rate than the mean longitudes alone. This leads to the appearance of small divisors during the averaging and we cannot remove the terms containing these combinations without obtaining a bad approximation. In fact, in $\mathcal H'_2$ it would come out an $\epsilon$ in the denominator, so that the real order would be one and not two. For this reason, we keep the terms containing the mentioned mean longitudes combinations (resonant terms) and the ones without mean longitudes (secular terms); the result of this choice is a semi-secular, or resonant, model. We truncate our expansion to the third order both for eccentricities and inclinations, therefore we have also to consider the terms of the resonance $4:1$ between Io and Ganymede.

As example, for the couple Io-Europa the perturbing function (up to the second order for simplicity) becomes
\begin{align*}
R^{(1,2)}=& -\frac{Gm_1m_2}{a_2}\Bigl[c_{(0,0,0,0,0,0)}(a_1/a_2) + c^1_{(0,0,0,0,0,0)}(e_1^2+e_2^2) \\
        & +c^2_{(0,0,0,0,0,0)}(s_1^2+s_2^2)+c_{(0,0,-1,1,0,0)} e_1e_2 \cos (\varpi_2-\varpi_1) \\
        & + c_{(0,0,0,0,-1,1)} s_1 s_2 \cos(\Omega_2-\Omega_1)\\
	& +c_{(-1,2,-1,0,0,0)}(a_1/a_2) e_1 \cos(2\lambda_2-\lambda_1-\varpi_1)\\
        & + c_{(-1,2,0,-1,0,0)}(a_1/a_2) e_2 \cos(2\lambda_2-\lambda_1-\varpi_2)\\
	& +c_{(-2,4,-2,0,0,0)} e_1^2 \cos (4\lambda_2-2\lambda_1-2\varpi_1)\\
        & + c_{(-2,4,0,-2,0,0)} e_2^2 \cos (4\lambda_2-2\lambda_1-2\varpi_2)\\
	& +c_{(-2,4,-1,-1,0,0)} e_1e_2 \cos (4\lambda_2-2\lambda_1-\varpi_1-\varpi_2)\\
	& +c_{(-2,4,0,0,-2,0)} s_1^2 \cos (4\lambda_2-2\lambda_1-2\Omega_1)\\
        & + c_{(-2,4,0,0,0,-2)} s_2^2 \cos (4\lambda_2-2\lambda_1-2\Omega_2)\\
	& +c_{(-2,4,0,0,-1,-1)} s_1s_2 \cos (4\lambda_2-2\lambda_1-\Omega_1-\Omega_2)\Bigr].
\end{align*}
The terms can be classified by means of their arguments:
\begin{align*}
& 0,\hspace{0.1cm}\varpi_2-\varpi_1,\hspace{0.1cm}\Omega_2-\Omega_1, & &\text{secular};\\
& 2\lambda_2-\lambda_1-\varpi_1,\hspace{0.1cm}2\lambda_2-\lambda_1-\varpi_2, & &\text{first order resonant};\\
& 4\lambda_2-2\lambda_1-2\varpi_2,\hspace{0.1cm}4\lambda_2-2\lambda_1-2\varpi_1, & & \\
& 4\lambda_2-2\lambda_1-\varpi_2-\varpi_1,\hspace{0.1cm}4\lambda_2-2\lambda_1-2\Omega_1, & & \\
& 4\lambda_2-2\lambda_1-2\Omega_2,\hspace{0.1cm}4\lambda_2-2\lambda_1-\Omega_2-\Omega_1, & &\text{second order resonant}.
\end{align*}
We compute the coefficients $c_{\mathbf j}$ at the beginning of the propagation and we leave them constant, except for the first one, which is free from eccentricities and inclinations, and the two first order terms. For them we do not neglect their dependence on the semi-major axes and their contribution in the Hamilton's equations.

Although we did not write them explicitly, we consider also $22$ third order terms plus $8$ that come out from the multiplication of the first order terms with second order factors. And this is just the perturbation between Io and Europa. Considering the perturbative function up to the third order will allow to improve the correspondence of the Io's eccentricity and the moons' inclinations with the ephemerides. Moreover, not keeping fixed the main terms' coefficients, we will obtain a better matching for the resonant angles.

In the case of the Sun's perturbation, the time scales are very different; in fact an entire revolution of Jupiter around the Sun takes almost $12$ years, which is comparable to the resonant time scale. Therefore, if we want a good representation of the resonant dynamics, we cannot eliminate all the terms with the longitude of the Sun. However, we can remove the terms containing the longitudes of the Galilean satellites, since in this case the angles are fast. Denoting with the subscript $s$ the Sun's orbital elements, the terms we keep (up to the second order) are the ones with arguments:
\begin{align*}
& 0,\hspace{0.1cm}\varpi_s-\varpi_i,\hspace{0.1cm}\Omega_s-\Omega_i, & &\text{secular};\\
&\lambda_s-\varpi_i,\hspace{0.1cm}2\lambda_s-2\varpi_i,\hspace{0.1cm}2\lambda_s-\varpi_s-\varpi_i, & &\\
& 2\lambda_s-2\Omega_i,\hspace{0.1cm}2\lambda_s-\Omega_s-\Omega_i, & &\text{mid-period}.
\end{align*}

Finally, for the zonal harmonics we consider the averaging of~\eqref{harm}. First of all we have to replace $\sin\phi_i$ and $\abs{\mathbf r_i}$ as functions of the orbital elements, and then we mean over the variable $\lambda_i$. We do not report here all the details, which can be found in~\cite{murder}, Chapter 6. We add to the Hamiltonian the resulting secular perturbative function:
\begin{align*}
R_J^{(i)}=-\frac{Gm_0}{a_i}\Bigl[&\frac{1}{2}J_2\Bigl(\frac{R_0}{a_i}\Bigr)^2-\frac{3}{8}J_4\Bigl(\frac{R_0}{a_i}\Bigr)^4 +\Bigl(\frac{3}{4}J_2\Bigl(\frac{R_0}{a_i}\Bigr)^2-\frac{15}{8}J_4\Bigl(\frac{R_0}{a_i}\Bigr)^4\Bigr)e_i^2 \\
    & -\Bigl(\frac{3}{4}J_2\Bigl(\frac{R_0}{a_i}\Bigr)^2-\frac{15}{8}J_4\Bigl(\frac{R_0}{a_i}\Bigr)^4\Bigr)s_i^2\Bigr].
\end{align*}

It is worth noting that, while the contribution of $J_4$ is quite limited, the secular effect of $J_2$ is the strongest perturbation in the system. Therefore, also for a basic model of the moons' dynamics, it must be taken into account.

After the averaging, we have that the new Hamiltonian is:
\[\mathcal H=\mathcal H_0+\epsilon\Bigl(\sum_i \tilde R_J^{(i)}+\sum_{i,k}\tilde R^{(i,k)}+\sum_{i}\tilde R^{(i,s)}\Bigr),\]
where $\mathcal H_0$ consists of the Keplerian terms of all the Galilean satellites, $\epsilon$ is the small parameter and the rest is the averaged perturbation. Resonant terms are present only in the mutual perturbations of the couples Io-Europa and Europa-Ganymede (first order) and Io-Ganymede (third order), while the others are secular terms (without mean longitudes), except for a few mid-period terms due to the interaction with the Sun.

We note that some of the coefficients $c_{\mathbf j}$ in~\eqref{3body} have both a direct and an indirect part. The expression of the second one changes if we consider an internal or an external perturber. However, for the couples satellite-Sun we consider an external perturbation, since the star is not affected by the moons. Instead, for the mutual perturbation between the satellites, there are few coefficients with an indirect part and for all of them the two versions coincide in value in the case of mean motion resonance $2:1$. For example, for the coefficient of the resonant term with argument $2\lambda_k-\lambda_i-\varpi_k$, denoting $\alpha=a_i/a_k$, the indirect part is $-2\alpha$ when we consider the perturbation acting on the internal satellite, and it is $-1/(2\alpha^2)$ in the other case. For bodies close to $2:1$ resonance $\alpha$ is almost $(1/2)^{(2/3)}\simeq0.62996$, therefore the two expressions are very near in value. We preferred the second version, since the term is more related to the outer satellite variables than to the inner. This choice resulted quite important to obtain a better representation of the resonant angles.

\section{Resonant variables}
A generic four-body problem has dimension $24$, that corresponds to six orbital elements per body. The averaging removes most of the terms of the Hamiltonian, so it is possible that some variables can be eliminated. For example $\lambda_4$ does not appear in the function, since all the terms with the longitude of Callisto have short period.

We perform a change of variables:
\begin{align*}
&\sigma_1=2\lambda_2-\lambda_1-\varpi_1, & \sigma_2&=2\lambda_2-\lambda_1-\varpi_2,\\
&\sigma_3=2\lambda_3-\lambda_2-\varpi_3, & \sigma_4&= -\varpi_4,\\
&\xi_1=2\lambda_2-\lambda_1-\Omega_1, & \xi_2&=2\lambda_2-\lambda_1-\Omega_2,\\
&\xi_3=2\lambda_3-\lambda_2-\Omega_3, & \xi_4&=-\Omega_4,\\
&\gamma_1=\lambda_1-3\lambda_2+2\lambda_3, & \gamma_2&=\lambda_2-2\lambda_3,\\
&\gamma_3=\lambda_3, & \gamma_4&=\lambda_4.
\end{align*}
We can write a linear generating function in order to obtain the conjugated momenta, which are $\Sigma_i=G_i$, $\Xi_i=H_i$ and $\Gamma_i$, where
\begin{align*}
&\Gamma_1=L_1+G_1+G_2+H_1+H_2\\
&\Gamma_2=3L_1+L_2+G_1+G_2+G_3+H_1+H_2+H_3\\
&\Gamma_3=4L_1+2L_2+L_3 \\
&\Gamma_4=L_4
\end{align*}

In the Hamiltonian both $\gamma_3$ and $\gamma_4$ are missing. Then their conjugated momenta remain constant along the motion and we can remove them. The number of variables decreases to $20$ and they comprehend only slow angles (with respect to the orbital periods). We substitute $e_i$ and $s_i$ ($i=1,4$) using the new momenta, taking into account the approximations
\[e_i=\sqrt{\frac{2\Sigma_i}{L_i}}, \qquad s_i=\sqrt{\frac{\Xi_i}{2L_i}}.\]
Since their contribution is quite limited, for all the terms of the mutual perturbations $R^{(i,k)}$ we keep constant the variables $L_i$, except for the ones such that also the coefficient $c_{\mathbf j}$ is not considered constant.

It is known that the angle $\sigma_1$ librates around $0$ and $\sigma_2$ and $\gamma_1$ librate around $\pi$. The frequency of their libration characterize the evolution of all the Galilean satellites' orbital elements: $\sigma_1$ and $\sigma_2$ are relative to the resonance between Io and Europa, $\sigma_2+\gamma_1$ to the resonance between Europa and Ganymede, $\gamma_1$ to the three-body resonance, involving the longitudes of all the three moons.

\begin{figure}[t]
\centering
{\epsfig{figure=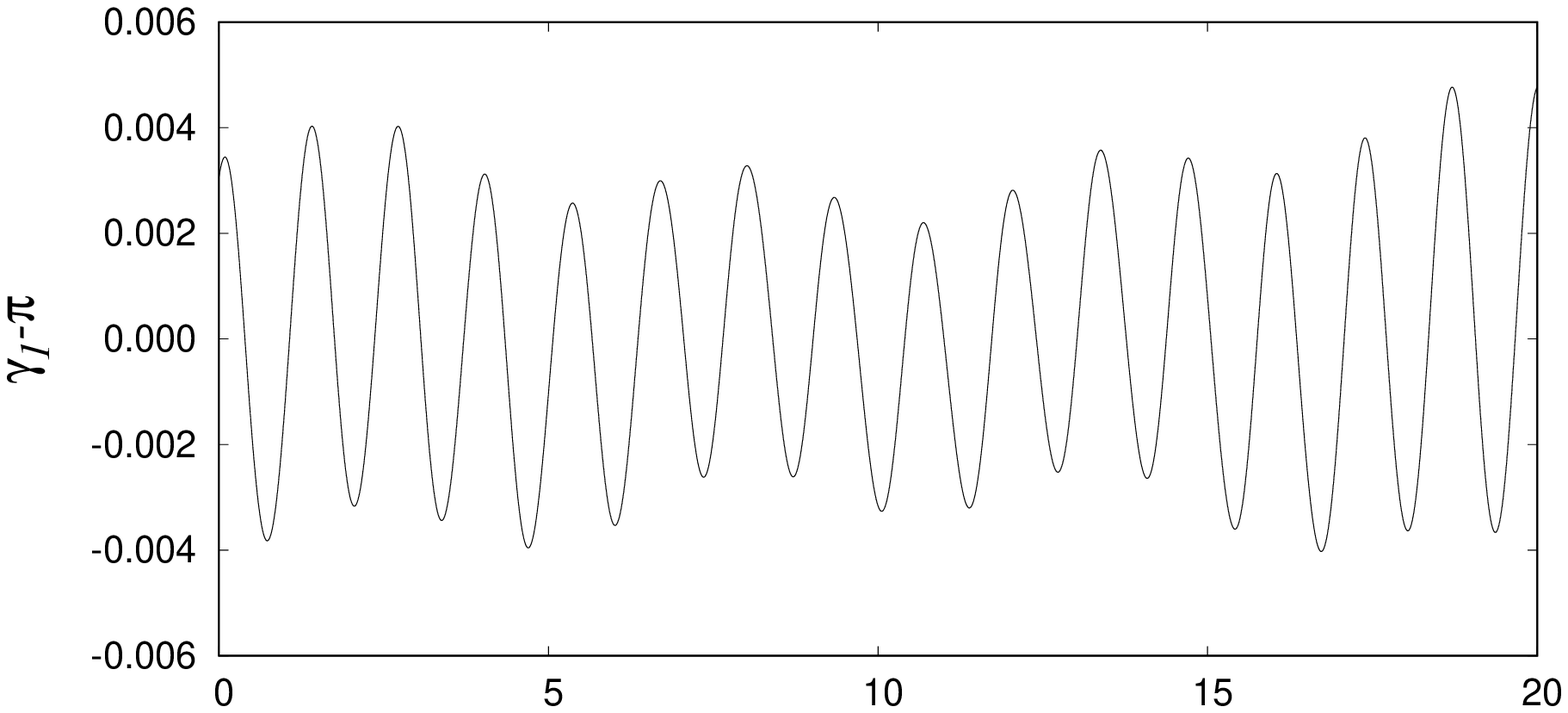,scale=0.3}} {\epsfig{figure=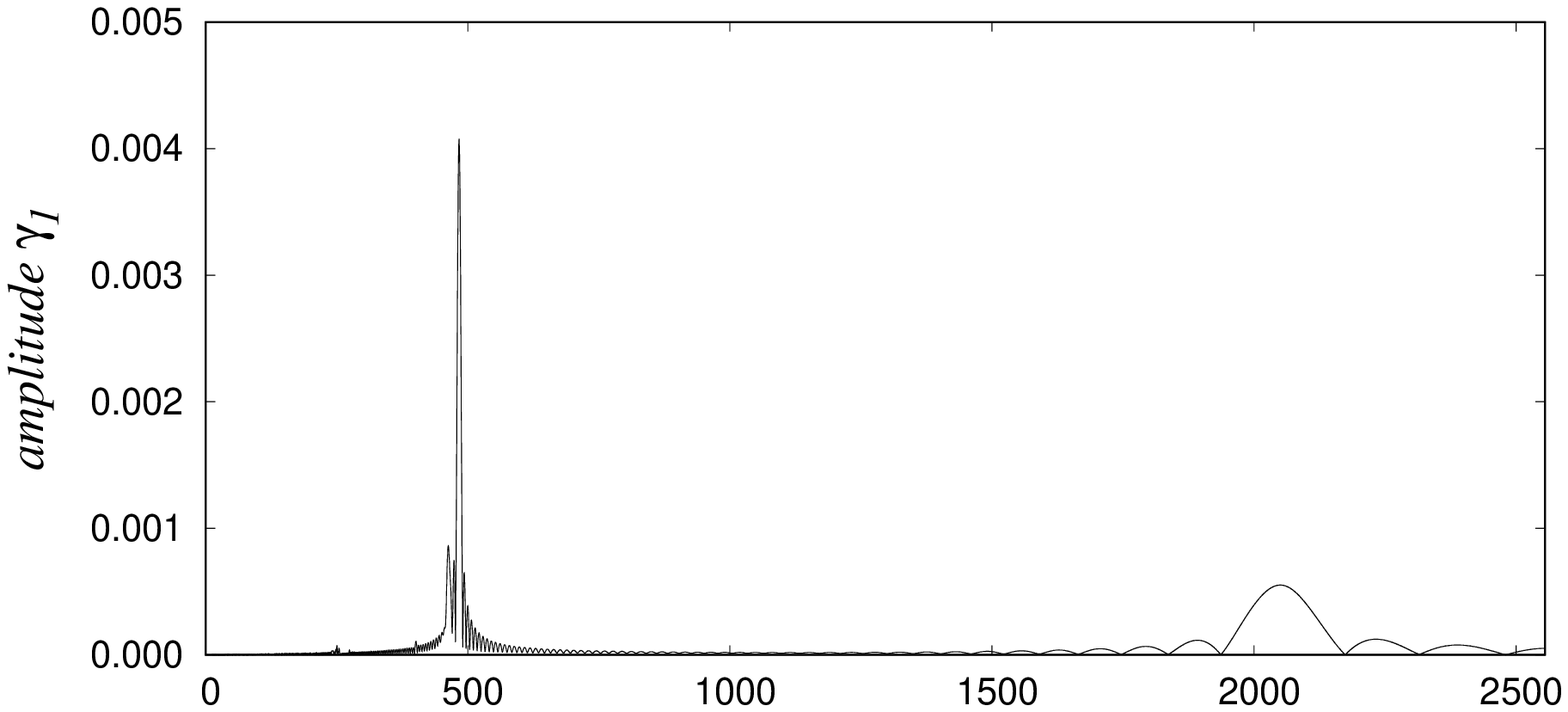,scale=0.3}}\\
{\epsfig{figure=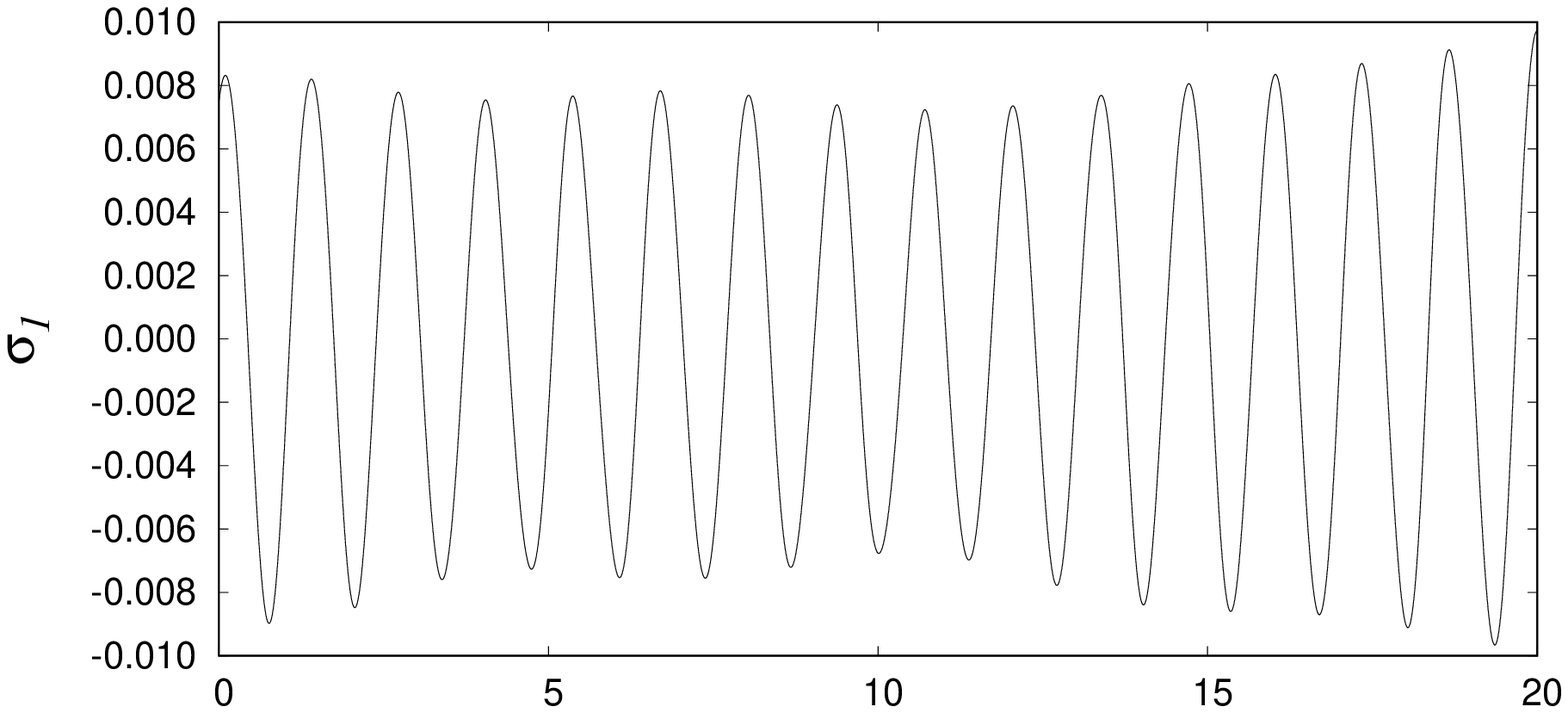,scale=0.3}} {\epsfig{figure=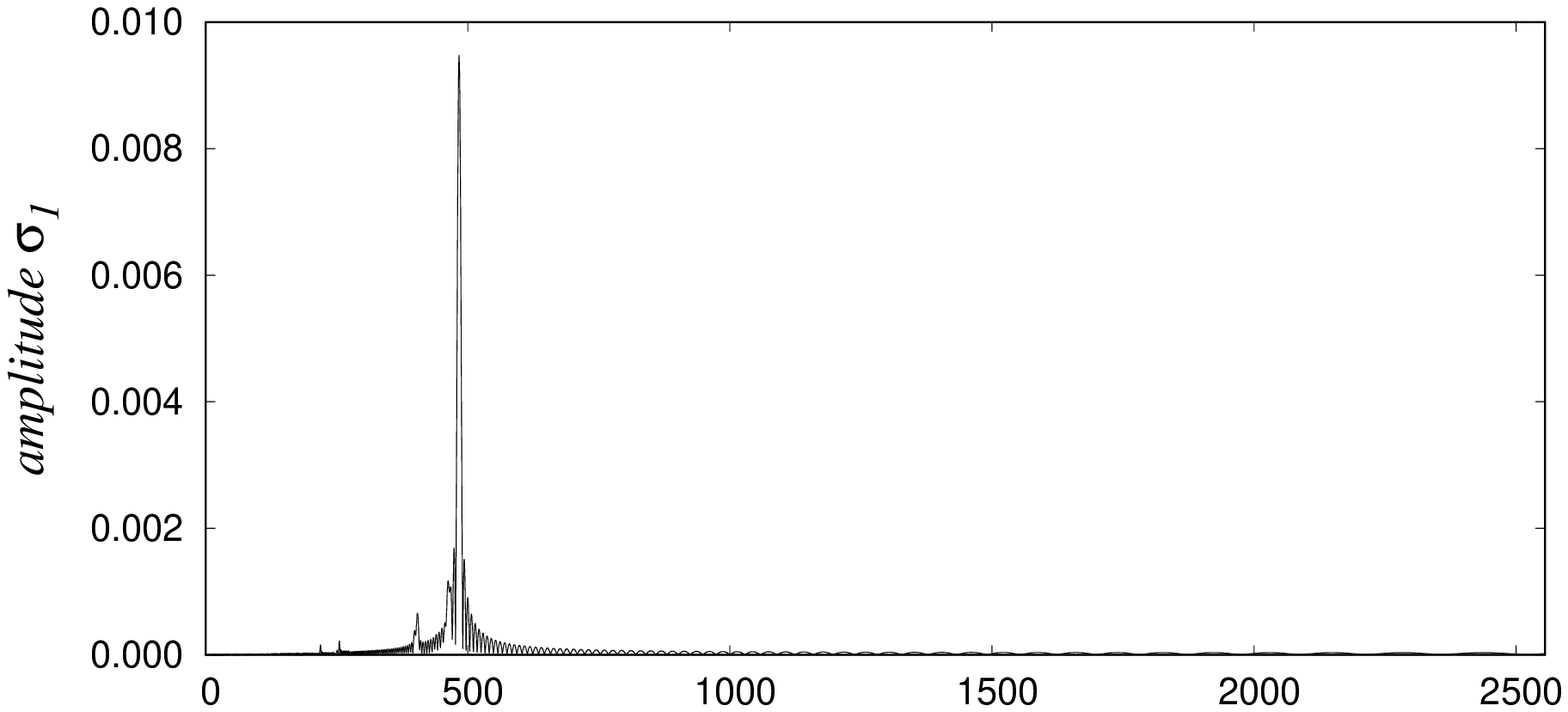,scale=0.3}}\\
{\epsfig{figure=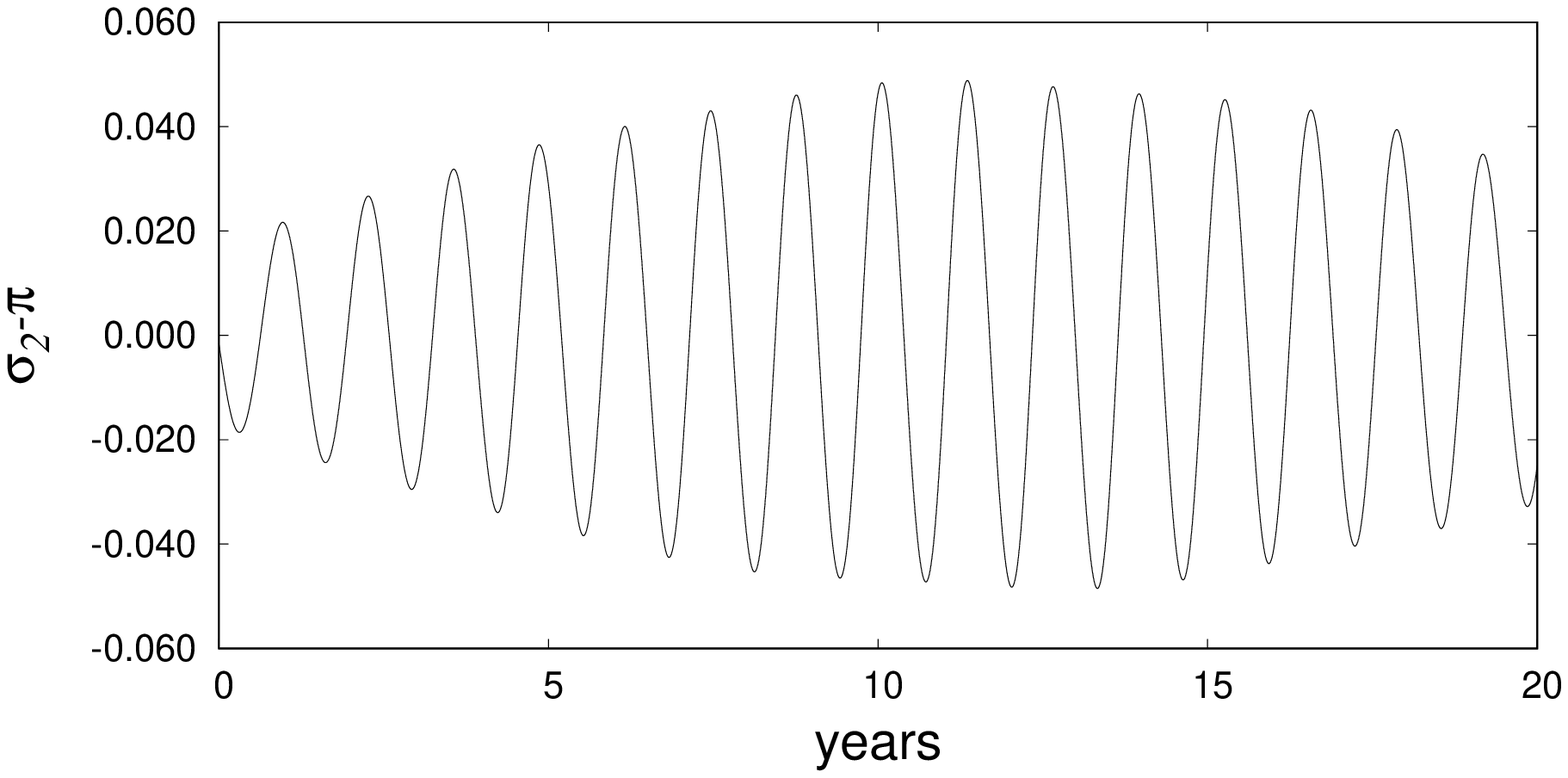,scale=0.3}} {\epsfig{figure=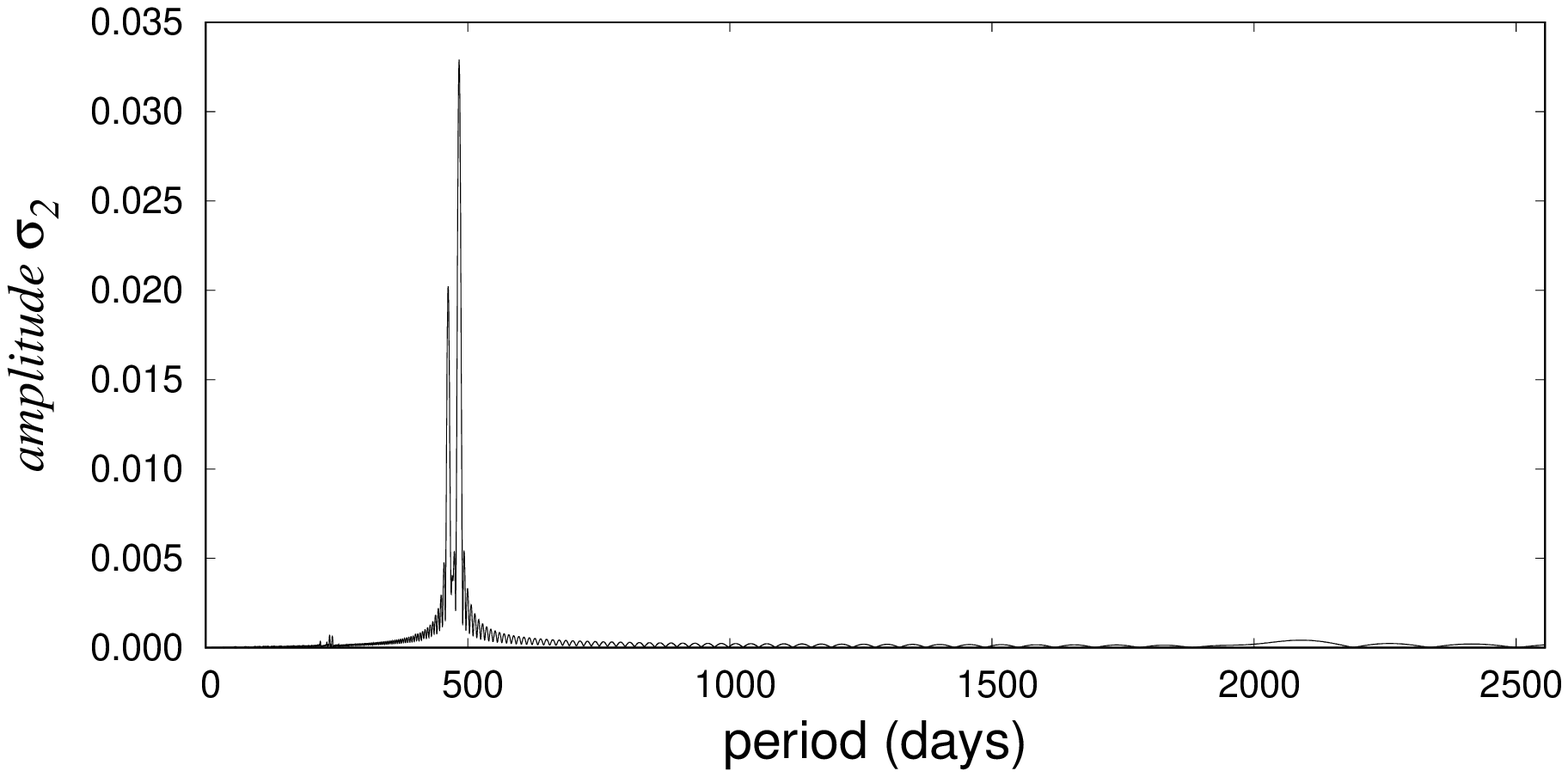,scale=0.3}}
\caption{The three resonant angles $\gamma_1$, $\sigma_1$ and $\sigma_2$ (left) and their spectra (right) from the propagation of the semi-analytical model. The angles $\gamma_1$ and $\sigma_2$ are shifted of $\pi$ radians in order to make clear the amplitude of their libration.}
\label{fig:resfreq}
\end{figure}

In order to simplify the Hamiltonian, we can pass to rectangular variables. The new momenta $x_i$, $u_i$ and the coordinates $y_i$, $v_i$ are defined by
\begin{align*}
x_i&=\sqrt{2\Sigma_i}\cos(\sigma_i), \hspace{1cm} y_i=\sqrt{2\Sigma_i}\sin(\sigma_i),\\
u_i&=\sqrt{2\Xi_i}\cos(\xi_i), \hspace{1.05cm} v_i=\sqrt{2\Xi_i}\sin(\xi_i).
\end{align*}
We do not change the variables $\Gamma_i$ e $\gamma_i$. In this way we treat polynomials of $x_i$ and $y_i$ (respectively $u_i$ and $v_i$), instead of $\cos(\sigma_i)$ and $\sin(\sigma_i)$ (respectively $\cos(\xi_i)$ and $\sin(\xi_i)$). In Appendix~\ref{completeham} we reported the entire writing of the Hamiltonian. From its expression we can compute the Hamilton's equations~\eqref{hamequ} which describe the motion of the four Galilean satellites. Once we have recovered suitable initial conditions for the semi-secular dynamics, we can solve numerically the associated Cauchy problem.

\section{Equations of motion and their integration}
The equations of motion are given by~\eqref{hamequ}. As the amount of the derivatives is quite large, we can appreciate the advantage to have polynomials instead of goniometric functions. However, we use a symbolic manipulation software in order to avoid undesired errors. Moreover, we have a natural error check in the conservation of the Hamiltonian, which must remain constant along the motion.

\begin{figure}[t]
\begin{center}
\epsfig{figure=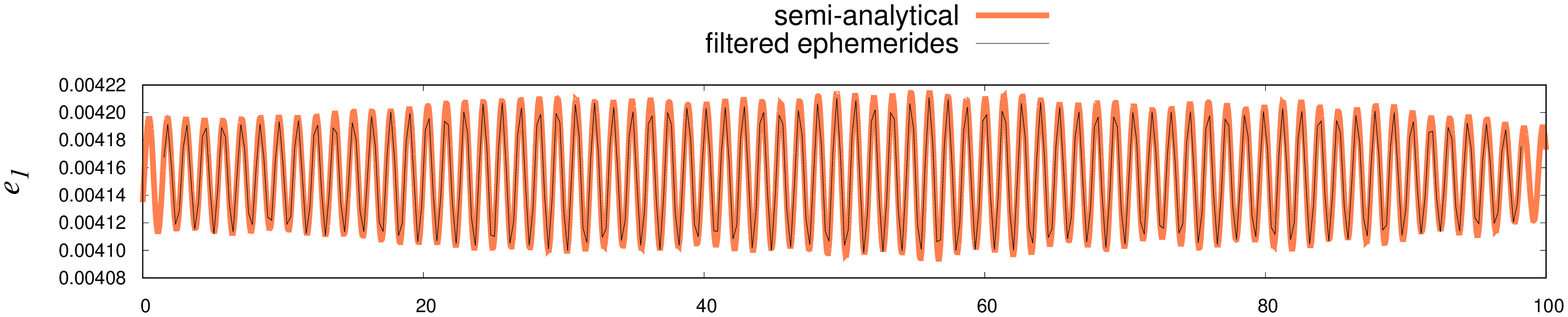,scale=0.39}\\
\epsfig{figure=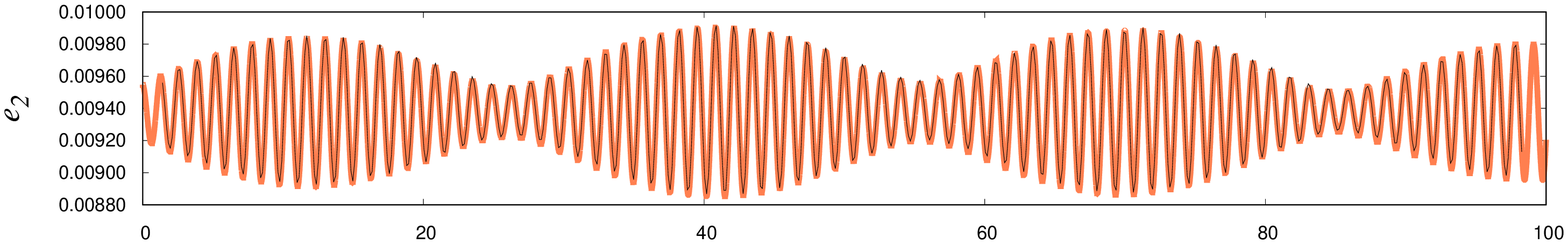,scale=0.39}\\
\epsfig{figure=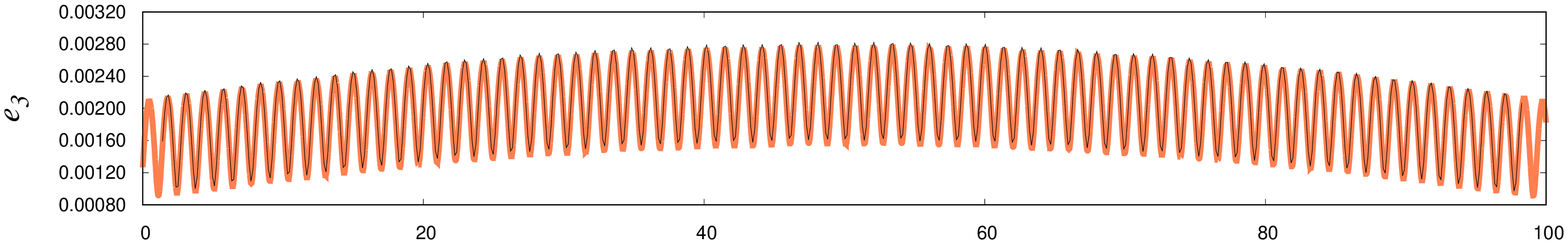,scale=0.39}\\
\epsfig{figure=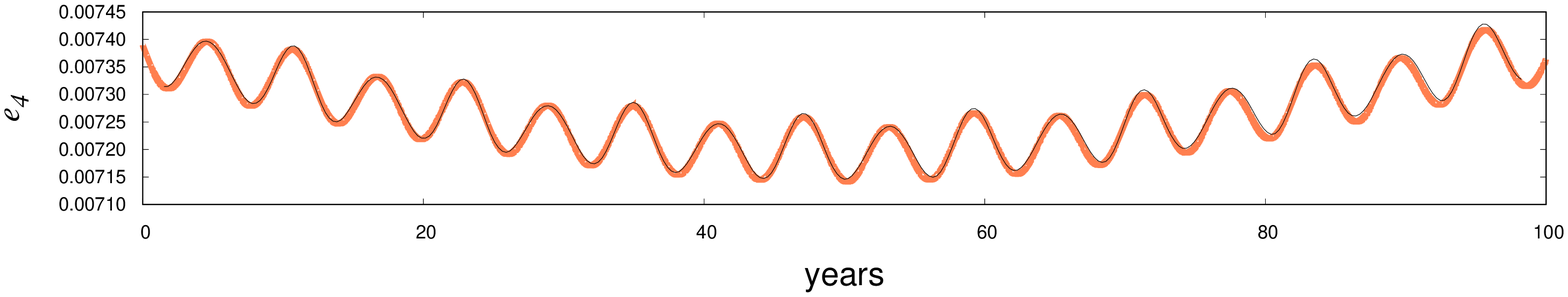,scale=0.39}
\end{center}
\caption{Galilean satellites' eccentricities over 100 years departing from J2000. In black we have the filtered elements taken from JUP310 ephemerides, while the thick coral line is the output of the semi-analytical model.}\label{fig:ecc100}
\end{figure}

Because of the complexity of the system, we integrate it numerically. For this aim we use an implicit three-stage Runge-Kutta-Gauss method, which is symplectic: in this way each step of integration is also a canonical transformation and we remain formally in a Hamiltonian context. Since the fast angles are removed, we can use a bigger time step. We choose a step of two days, that is small enough to obtain a precise propagation.

We cannot take the initial conditions from the ephemerides without any manipulation, because they describe the osculating elements. They are the instantaneous orbital elements and they suffer also the short period effects. If we took them as initial conditions, we could risk to get too much high or too much low values, because of the wide high frequencies oscillations. Therefore, we apply a digital filter (using the code giffv included in the orbit9 software, aimed to the numerical propagation of celestial bodies and described in~\cite{orbfit}, and based on the theory presented in~\cite{filt}), in order to remove the high frequencies from the moons' ephemerides. In particular, in this article we chose the JPL ephemerides JUP310 (\cite{jup310}). After this procedure we have a sort of synthetic mean elements, that we can use as initial conditions for the resonant semi-analytical model. The term synthetic is taken from the theory presented in~\cite{synthetic} and it is used as opposed to analytic, in order to mean that these elements are obtained only through numerical procedures.

From the filtered ephemerides series we take the initial conditions at J2000 and we perform a $100$ years integration. First of all, we present the output for the resonant angles of the system: in \figurename~\ref{fig:resfreq} we reported the evolution of $\sigma_1$, $\sigma_2$ and $\gamma_1$ and at their side the corresponding spectra. In order to highlight the libration periods, we limited the time interval to $20$ years.

\begin{figure}[t]
\begin{center}
\epsfig{figure=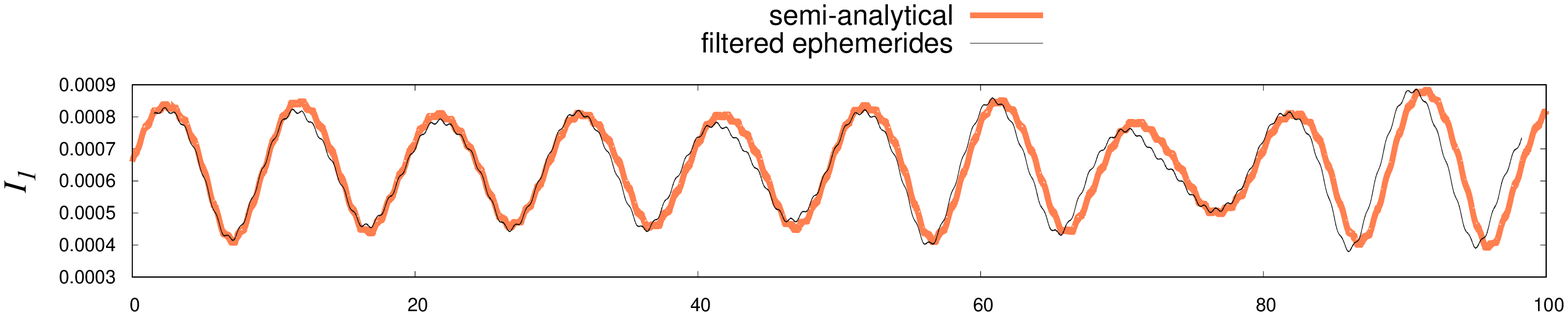,scale=0.39}\\
\epsfig{figure=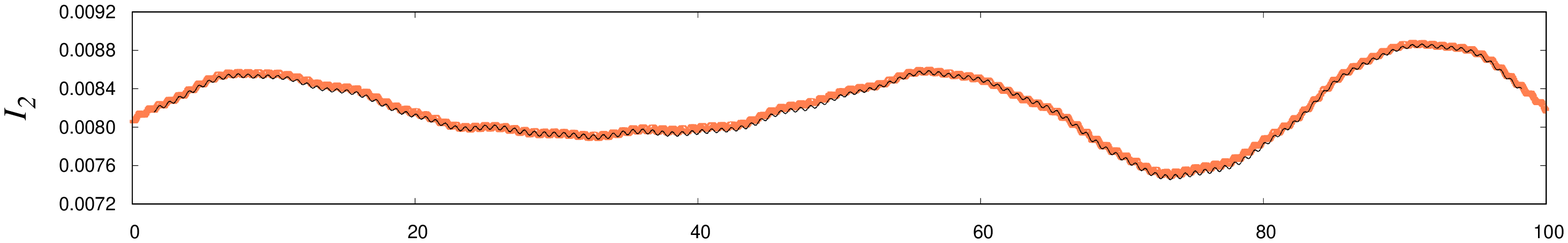,scale=0.39}\\
\epsfig{figure=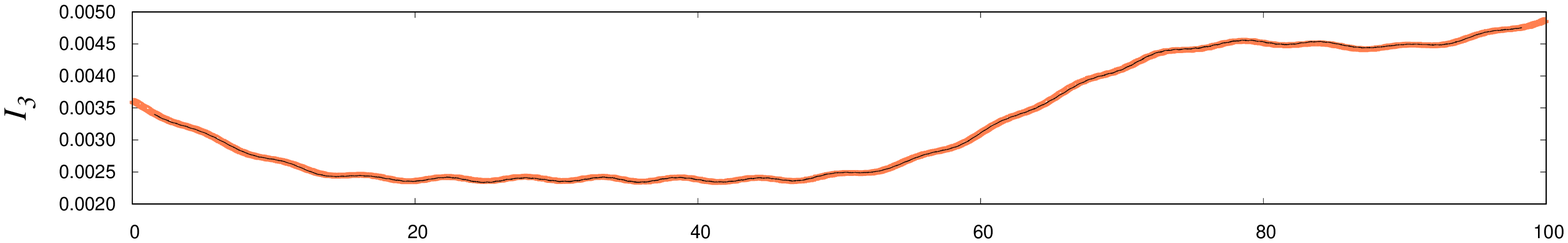,scale=0.39}\\
\epsfig{figure=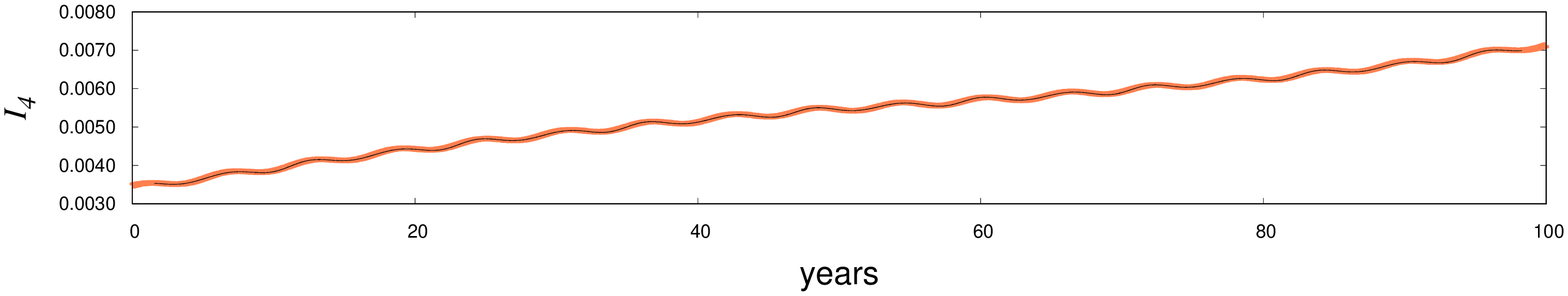,scale=0.39}
\end{center}
\caption{Galilean satellites' inclinations over 100 years departing from J2000. In black we have the filtered elements taken from JUP310 ephemerides, while the thick coral line is the output of the semi-analytical model.}\label{fig:inc100}
\end{figure}

Apart from the period of $486.89$ days, which is the circulation time of $\gamma_2$ and it is related to the inequality
\begin{equation}
\label{ineq}
\nu=n_1-2n_2=n_2-2n_3>0,
\end{equation}
each resonant angle has its characteristic period, which is associated to its libration. For $\sigma_1$ we found a period of almost $403.82$ days, for $\sigma_2$ almost $462.45$ days and for $\gamma_1$ almost $2047.85$ days. These values are in good agreement with the ephemerides and previous studies of the Galilean satellites' dynamics. In particular, from~\cite{laineyfreq}, which presents an exhaustive frequency analysis of a complete numerical model of the moons, they are respectively $486.81$, $403.52$, $462.51$ and $2059.62$.

\begin{figure}[t]
\begin{center}
\epsfig{figure=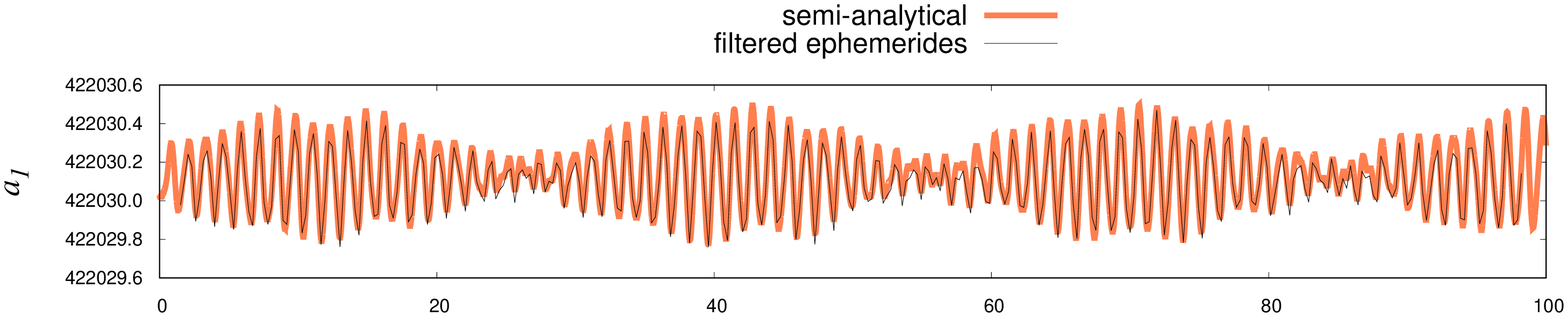,scale=0.39}\\
\epsfig{figure=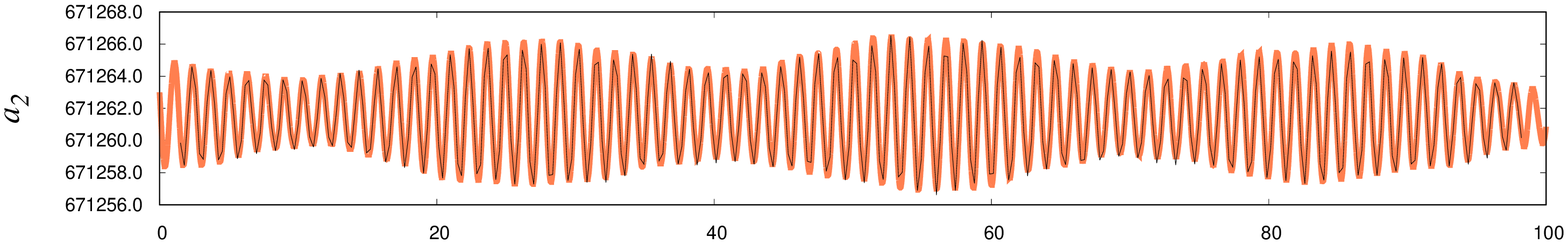,scale=0.39}\\
\epsfig{figure=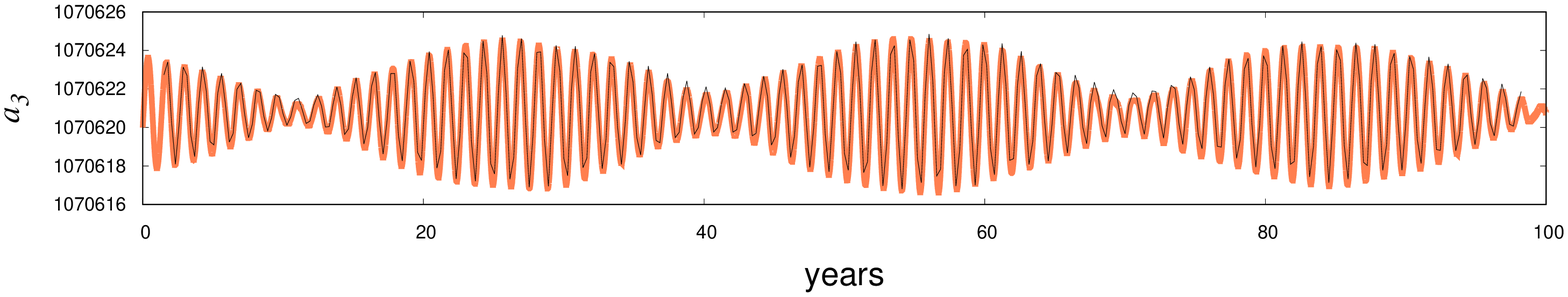,scale=0.39}
\end{center}
\caption{Galilean satellites' semiaxes over 100 years departing from J2000. In black we have the filtered elements taken from JUP310 ephemerides, while the thick coral line is the output of the semi-analytical model. We do not report the semi-axis of Callisto, since it is constant in the semi-secular model.}\label{fig:semiaxe100}
\end{figure}

\begin{figure}[t]
\begin{center}
\epsfig{figure=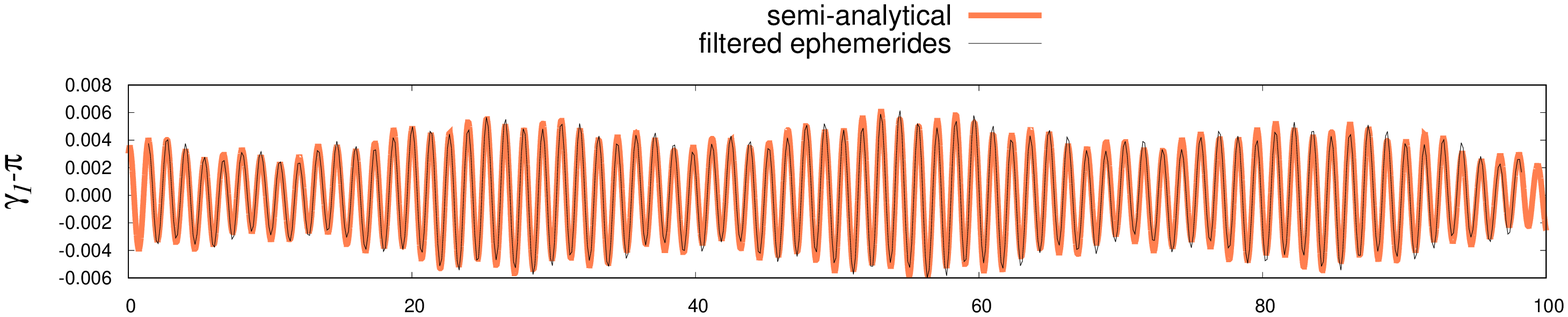,scale=0.39}\\
\epsfig{figure=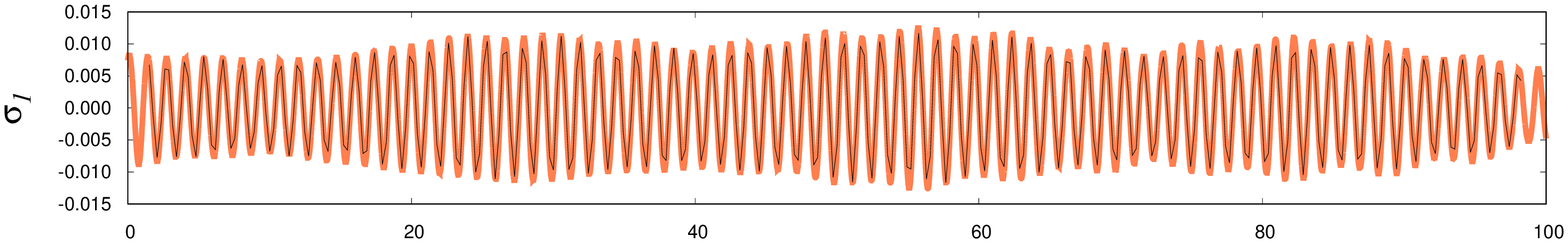,scale=0.39}\\
\epsfig{figure=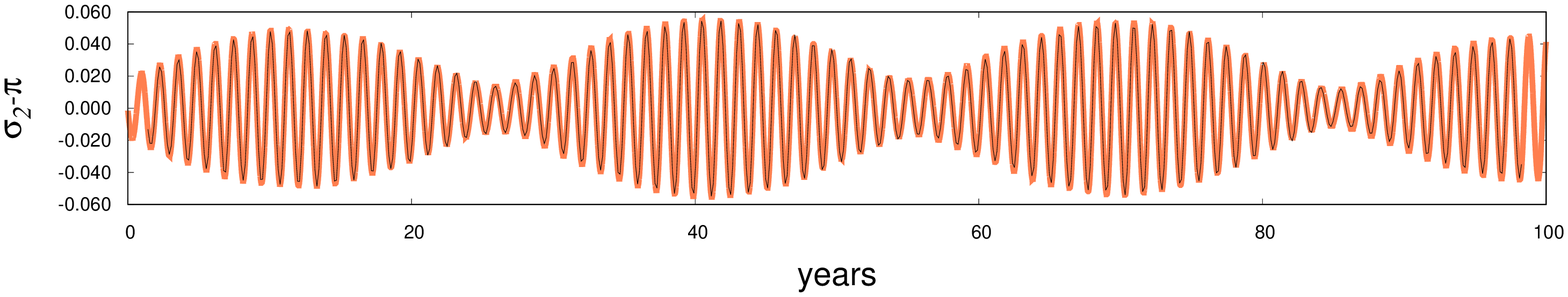,scale=0.39}
\end{center}
\caption{The three resonant angles $\gamma_1$, $\sigma_1$ and $\sigma_2$ over 100 years departing from J2000. In black we have the filtered elements taken from JUP310 ephemerides, while the thick coral line is the output of the semi-analytical model. The angles $\gamma_1$ and $\sigma_2$ are shifted of $\pi$ radians in order to make clear the amplitude of their libration.}\label{fig:res100}
\end{figure}

In \figurename~\ref{fig:ecc100},~\ref{fig:inc100},~\ref{fig:semiaxe100} and~\ref{fig:res100} we reported some of the satellites' mean orbital elements, obtained integrating the semi-analytical model. In the same pictures we plotted the filtered series of the ephemerides, in order to compare them with our model. From the figures we can see that the two output are almost coincident: all the computed mean elements correspond qualitatively well to the ephemerides. In order to quantify the differences between the two models (we indicate with $x$ the output of the semi-analytical model and with $x_e$ the filtered ephemerides), we compute the relative standard deviation of the residuals
\[\sigma^R=\frac{STD(x-x_e)}{STD(x_e)}.\]
In \tablename~\ref{tab:residual} we reported the results for the main orbital elements: on average the variations with respect to the mean value differ of about $10\%$ for a propagation of one century, apart from the inclinations, for which we find a discordance of about $34\%$ for Io and almost $1\%$ for the other moons. The high value for the Io's inclination is due to the fast rate of change of $\Omega_1$, which amplifies the differences of the two models after few tens of years. We chose a total time of $100$ years so that it is possible to appreciate both the resonant and the secular frequencies. For example, in the Io and Europa's eccentricities and all the semi-major axes the signals due to the variation of the resonant angles are predominant, especially the beat of the frequencies of $\sigma_2$ and $\gamma_2$; while in the outer satellites' eccentricities and all the inclinations the secular effects are much more appreciable.

In particular, it is evident the same secular frequency in the Ganymede and Callisto's eccentricity: with the semi-analytical model it is easy to prove that it is due to the term $e_3e_4\cos(\varpi_4-\varpi_3)$ of the mutual perturbation between the two moons. In fact, the period of the signal is $181.5$ years, which is very near to the circulation time of $\varpi_4-\varpi_3$ (almost $182.2$ years with a linear fit of a longer propagation). This is a clear example of the importance to consider Callisto in the model. Moreover, the couple Ganymede-Callisto has very interesting characteristics, as some combinations of the angles (in particular $7\lambda_4-3\lambda_3$ and $4\varpi_4-\varpi_3$) are almost resonant, and in the future it could be worth to study their effects on the dynamics.

Since JUP310 ephemerides cover the years between 1900 and 2100, we cannot use them for validating our model for longer time intervals. However, we can refer to works in literature in which the dynamics of the Galilean satellites is propagated for thousands of years. For example, in~\cite{musotto} the authors used a numerical method based on the Wisdom-Holman theory (\cite{holman}), and presented the output of a $2000$ years propagation. This method consists in activating the perturbations just at certain instants evenly spaced and using the Keplerian dynamics in the rest of the time. Their dynamical model differs from ours just for the adding of the other giant planets of the Solar System, which are important mainly for the evolution of the Jupiter's orbit. We performed a propagation of the same duration and we compared the output with the numerical series presented in~\cite{musotto}, which were previously filtered in order to remove short periods and part of the resonant ones. In the paper's results the authors focused on the long period signals found in their integration, such as the one of $181$ years present in the spectrum of the Ganymede and Callisto's eccentricities we already mentioned. Moreover, they obtained a main signal of $140$ years in the Ganymede's inclination and of $550$ years in Callisto's inclination. All these values are in good agreement with the ones we found and we reported in \tablename~\ref{tab:sec_per}. In particular, they are related to the circulation of $\Omega_3$ and the long period oscillation of $\Omega_4$ respectively, which are strongly influenced by the Sun.

\begin{table}[t!]
{\small
\begin{center}
\caption{Relative standard deviation $\sigma^R$ of the orbital elements' residuals over $100$ years.}\label{tab:residual}
\noindent\begin{tabular}{lrrrrrrr}
\toprule
& $a_1$ & $a_2$ & $a_3$ & $e_1$ & $e_2$ & $e_3$ & $e_4$ \\
\midrule
$\sigma^R$ & $0.13$ & $0.10$ & $0.09$ & $0.08$ & $0.07$ & $0.07$ & $0.08$   \\
\midrule
\midrule
& $I_1$ & $I_2$ & $I_3$ & $I_4$ & $\gamma_1$ & $\sigma_1$ & $\sigma_2$ \\
\midrule
$\sigma^R$ & $0.34$ & $0.03$ & $0.01$ & $0.01$ & $0.16$ & $0.09$ & $0.07$ \\
\bottomrule
  \end{tabular}
\end{center}
}
\end{table}

Returning to the resonant dynamics, from \figurename~\ref{fig:res100} and \tablename~\ref{tab:residual} we can appreciate how we obtain very good matches between the resonant angles, that allows not to find undesired signals in the output. A non-negligible difference in the amplitudes and the frequencies of the angles' librations generally leads to a degradation in the comparison, in particular for eccentricities and semi-major axes. In this sense $\gamma_1$ is the most challenging angle. In fact, it is very sensitive to changes in the dynamics or in the initial conditions and a variation in its amplitude can lead to a bad approximation of the actual resonant dynamics. For this reason we had to pay attention to small details in the model, such as considering the variation of the coefficients $c_{\mathbf j}$ of the main terms and choosing the internal perturber version for the indirect part of the terms with arguments $2\lambda_k-\lambda_i-\varpi_k$. Moreover, as we said, the correspondence for $\sigma_1$ and $e_1$ is improved thanks to the adding of the third order terms.

It is worth noting that we can recover the short period dynamics using the inverse of the canonical transformation we introduced to define the secular Hamiltonian~\eqref{newh}. If we want the osculating elements at a certain time, we can integrate the semi-secular model up to that time, return to the initial set of variables and then apply the inverse of the transformation to the secular coordinates. In this way, apart from describing the resonant and secular dynamics, the model can provide ephemerides of the moons.

\section{Tidal dissipation}
Tides cause significant effects on the dynamics. In particular, the tides between Io and Jupiter change in a significant way the energy of the system, leading to a migration of the Galilean satellites' orbits. In~\cite{kaula} it was presented a first detailed analysis of the tidal dissipation, through an expansion of the potential and the introduction of the lag angles. Before adding the dissipation into the equations, we give a short description of the phenomena we take into account.

\begin{table}[t]
{\small
\begin{center}
\caption{Rates of change of the pericenters and nodes' longitudes and main secular periods of the eccentricities and inclinations, computed from a $2000$ years propagation of the semi-analytical model. ($^+$) For $\Omega_4$ we did not find an evident linear rate, but an oscillation with a very long period of almost $580$ years. (*) We did not write any value for $e_1$ and $e_2$, since they are characterized by resonant effects mainly. For the inclination of Europa we reported two values, since the contribution of the two signals is almost equivalent.}\label{tab:sec_per}
\noindent\begin{tabular}{lrrrrrrrr}
\toprule
& $\dot\varpi_1$ & $\dot\varpi_2$ & $\dot\varpi_3$ & $\dot\varpi_4$ & $\dot\Omega_1$ & $\dot\Omega_2$ & $\dot\Omega_3$ & $\dot\Omega_4$\\
\midrule
Linear rate & -$4.710$ & -$4.710$ & $0.046$ & $0.012$ & -$0.841$ & -$0.209$ & -$0.046$ & $^+$ \\
\midrule
\midrule
 & $e_1$ & $e_2$ & $e_3$ & $e_4$ & $I_1$ & $I_2$ & $I_3$ & $I_4$\\
\midrule
Main period &  * & * & $181.52$ & $181.67$ & $9.91$ & $30.19$ & $137.44$ & $580.06$ \\
            &    &   &          &          &        & $38.63$ &          &          \\
\bottomrule
  \end{tabular}
\end{center}
}
\end{table}

The tides that a satellite raises on its planet are not aligned perfectly with the conjunction line between the two bodies, but they are dragged by the rotation of the planet. In the case of Jupiter and Io, the spin of the planet is faster than the mean motion of the moon, therefore the maximum tides are pushed ahead, forming an angle $\delta$ with the conjunction line of the bodies. Consequently, Io acts a torque on the two tidal bulges of Jupiter, which slows the spin of the planet. For the conservation of the system's angular momentum, the semi-major axis of Io must increase, therefore the moon moves away from Jupiter and decelerates.

Also the tides that a planet raises on its moon change its orbit. If we consider a corotating satellite with eccentricity different from zero (such as Io), the point of maximum tides moves both radially and transversally on the surface. In fact, the distance of the moon from the planet is not constant, and then tides height changes during an orbital cycle (radial tides). Moreover, the satellite looks at the empty focus of its orbit, so that the point of maximum tides oscillates around a fixed point on the surface of the moon (librational tides). This continuous compression is the reason of the energy dissipation within the body, stealing energy directly from its orbit. In this case, the semi-major axis tends to decrease and consequently Io accelerates. Moreover, since only the energy is dissipated and the angular momentum remains constant, the eccentricity decreases.

If we want to add these effects, we have to include a dissipative term both on the semi-major axis and on the eccentricity of Io. From~\cite{malho} and~\cite{peale} we have the expressions for these contributions
\begin{align}
\label{dissa}
\frac{da_1}{dt}&=\frac{2}{3}c(1-7De_1^2)a_1, \\
\label{disse}
\frac{de_1}{dt}&=-\frac{7}{3}cDe_1.
\end{align}
The coefficients $c$ and $D$ are
\begin{gather*}
c=\frac{9}{2}\frac{k_2^0}{Q^0}\frac{m_1}{M_0}\Bigl(\frac{R_0}{a_1}\Bigr)^5n_1,\\
D=\frac{k_2^1}{k_2^0}\Bigl(\frac{R_1}{R_0}\Bigr)^5\Bigl(\frac{m_0}{m_1}\Bigr)^2\frac{Q^0}{Q^1}.
\end{gather*}
In the formula above $k_2$ is the Love number of order $2$, which is an adimensional parameter that describes the body deformation due to the tidal field, and $Q$ is the tidal quality factor, which is an adimensional parameter that quantifies the energy dissipation due to the tides. The superscript indicates if they are referred to Io or Jupiter. In the equations~\eqref{dissa} and~\eqref{disse} the term with $cD$ is due to the tides on Jupiter, while the term with $c$ alone is due to the tides on Io. In~\eqref{disse} we consider only the effect of the tides on the satellite, while we omit the contribution of the tides on the planet, since it is far smaller.

Following~\cite{malho}, we can translate the differential equations~\eqref{dissa} and~\eqref{disse} in differential equations for $L_1$ and $G_1$
\begin{align*}
&\dot G_1 = -\frac{14}{3}cDG_1,\\
&\dot L_1 = \frac{1}{3}cL_1-\frac{14}{3}cDG_1,
\end{align*}
and for the new momenta $\Gamma_i$ $(i=1,3)$ and $\Sigma_1$
\begin{align*}
&\dot \Sigma_1 = \dot G_1,\\
&\dot \Gamma_1 = \dot L_1 + \dot G_1,\\
&\dot \Gamma_2 = 3\dot L_1 + \dot G_1,\\
&\dot \Gamma_3 = 4\dot L_1.
\end{align*}

Although we are considering a dissipation, i.e. a non-conservative effect, we continue to use Hamilton's equations, as the tides are smaller than the other forces in the system. At the end of the propagation we can check if the amount of energy dissipated (given by $\dot E=-E(\dot a/a)$) coincides with the variation of the Hamiltonian (which contains the orbital energy of the moons).

\begin{figure}[t]
\centerline{\epsfig{figure=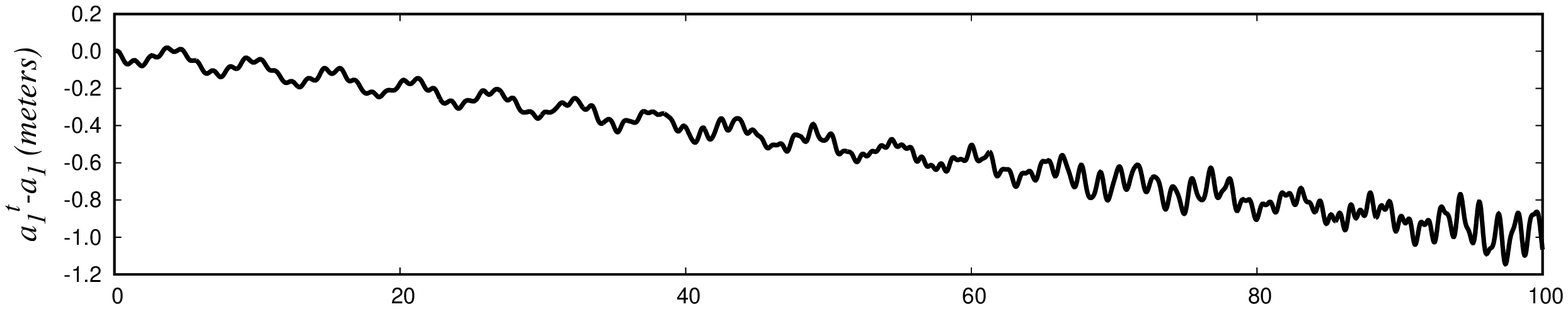,scale=0.45}}
\centerline{\epsfig{figure=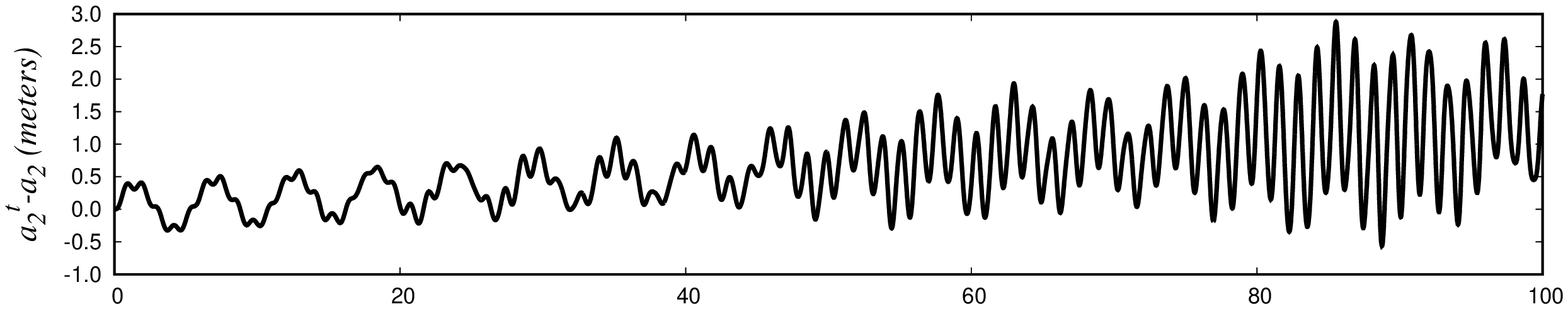,scale=0.45}}
\centerline{\epsfig{figure=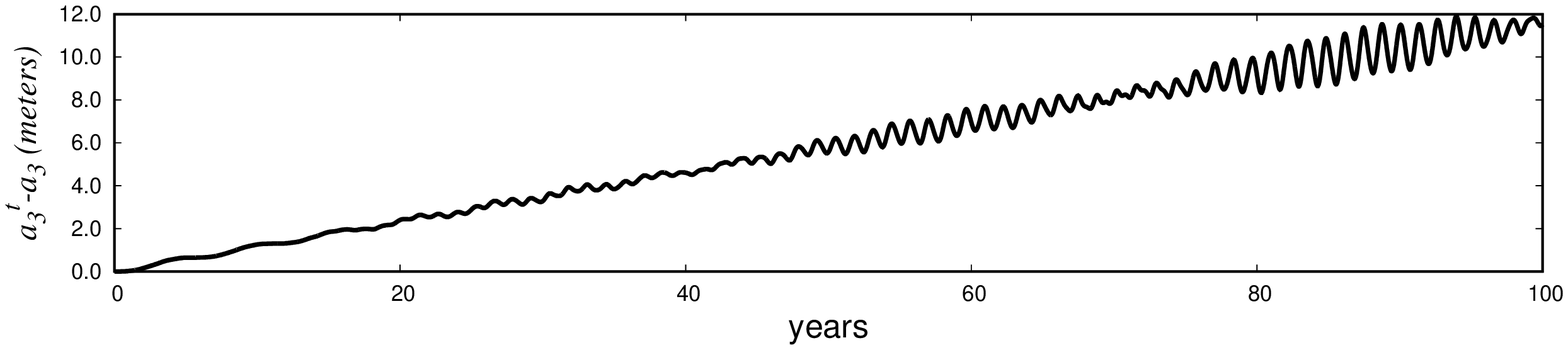,scale=0.45}}
\caption{Effect of the tidal dissipation on the Galilean satellites orbits, in term of the differences in the semi-major axes between a simulation with dissipation and another one without dissipation.}\label{fig:adiss}
\end{figure}

The resonant interaction spreads the dissipation in the system, affecting also Europa and Ganymede's orbits. We can perform a test with our semi-secular model in order to see if it captures well this aspect of the dynamics.

For the dissipative parameters $k_2^1/Q^1$ and $k_2^0/Q^0$ we consider the values reported in~\cite{laineynature}, which are $0.015$ and $1.102\times10^{-5}$ respectively. In \figurename~\ref{fig:adiss} we show the difference in the semi-major axes between a simulation with the dissipation and another one without the dissipation. It is evident a linear term in the semi-major axes of all the three Galilean satellites; after 100 years they change of about $1$-$10$ meters, that correspond in a shift along the orbit of almost tens-hundreds of kilometers.

The results we obtain in terms of variation in the mean motion over mean motion (units $10^{-10}$ $1/year$) are
\begin{equation}
\label{dotn}
\frac{\dot{n}_1}{n_1}=0.355, \hspace{1.5cm} \frac{\dot{n}_2}{n_2}=-0.303, \hspace{1.5cm}\frac{\dot{n}_3}{n_3}=-1.626
\end{equation}

In order to validate these results, we compute the same quantities using the corresponding numerical model, written in Cartesian coordinates and including only the forces we considered for the semi-analytical model. For its integration we take the initial conditions from the JUP310 ephemerides, without any operation of filtering; since we included the main perturbations in the system, the numerical model approximates well the moons' ephemerides. For simulating the tidal dissipation, we need a formula for the force that a body undergoes because of the tides it raises on another body. In~\cite{mignard80} it was presented the force acting on the Moon due to the tides it generates on the Earth, introducing a time delay in the position of the satellite. The same formulation was used for the numerical investigation of the tidal evolution of other satellites in~\cite{efro} and~\cite{laineymars}. For the development of a more accurate dynamical model of the Galilean satellites, in~\cite{laineynature} the authors generalized the formula of this force also in the case of tides raised on a satellite and the same was used for the dynamics of the Saturn's system (\cite{laineysaturn}). From these two articles we know that in both cases, planet or satellite's tides, the tidal force can be written as follows
\begin{equation}
\label{tidemignard}
\mathbf F=-3\frac{k_2Gm^2R^5}{r^7}\Delta t\Bigl(2\frac{\mathbf{r}}{r}\frac{\mathbf{r}\cdot\mathbf{v}}{r^2}+\frac{\mathbf r \times \boldsymbol{\omega} + \mathbf v}{r}\Bigr),
\end{equation}
where $m$, $\mathbf r$ and $\mathbf v$ are the mass, position and velocity of the body that raises the tides (we used the notation $r=\abs{\mathbf r}$), while $k_2$, $R$ and $\boldsymbol{\omega}$ are the Love number, radius and angular velocity of the other body. Although the formula is the same in both cases, the variables and the parameters must be chosen carefully. The time lag $\Delta t$ contains the quality factor $Q$ and it is defined as
\[\Delta t=\frac{T}{2\pi}\arctan\Bigl(\frac{1}{Q}\Bigr)\simeq \frac{T}{2\pi}\frac{1}{Q},\]
where $T$ is the period of one tidal cycle. Following~\cite{laineynature} and denoting with $n$ the mean motion of the satellite, in the case of tides on the planet $T=2\pi/(2(\abs{\boldsymbol{\omega}}-n))$, which is the time that the moon takes to pass from a point over the planet's surface to its antipode, so that the tidal deformation comes back to its initial configuration. Instead, for tides acting on a corotant satellite $T=2\pi/n$, which is the time that the tidal deformation takes to complete a whole libration.

While for Jupiter we can consider a constant spin rotation, we cannot do the same for Io, as we supposed a corotant state of the moon. In~\cite{murder}, Chapter 4, the authors expanded the tidal potential acting on a corotant satellite and obtained two main terms, one due to the radial oscillation of the tides and the other due to the librational tides. Assuming that the satellites' $Q$ is constant, the two tidal oscillations are linear and their contribution can be calculated separately. They computed the work related to the tidal deformation for both the terms and they found that the energy dissipated by the librational tides is exactly $4/3$ times the one of the radial tides. Therefore, for Io we adopt the following approach: first we evaluate the dissipation due to the radial tides considering $\boldsymbol{\omega}=\mathbf{r}\times\mathbf{v}/r^2$,
\begin{align*}
\mathbf F&=-3\frac{k_2Gm^2R^5}{r^7}\Delta t\Bigl(2\frac{\mathbf{r}}{r}\frac{\mathbf{r}\cdot\mathbf{v}}{r^2}+\frac{1}{r}\Bigl(\mathbf r \times \frac{\mathbf{r}\times\mathbf{v}}{r^2} + \mathbf v \Bigr) \Bigr)\\
&=-3\frac{k_2Gm^2R^5}{r^7}\Delta t\Bigl(2\frac{\mathbf{r}}{r}\frac{\mathbf{r}\cdot\mathbf{v}}{r^2}+\frac{1}{r}\Bigl( \frac{\mathbf r \cdot (\mathbf{r}\cdot\mathbf{v})-\mathbf v \cdot (\mathbf{r}\cdot\mathbf{r})}{r^2} + \mathbf v \Bigr) \Bigr)\\
&=-3\frac{k_2Gm^2R^5}{r^7}\Delta t\Bigl(3\frac{\mathbf{r}}{r}\frac{\mathbf{r}\cdot\mathbf{v}}{r^2}\Bigr)
\end{align*}
and then we add $4/3$ of their effect for taking into account the contribution of the \text{librational} tides. In the end we have a factor $7$ inside the brackets, instead of $3$
\begin{equation}
\label{Fdiss}
\mathbf F=-3\frac{k_2Gm^2R^5}{r^7}\Delta t\Bigl(7\frac{\mathbf{r}}{r}\frac{\mathbf{r}\cdot\mathbf{v}}{r^2}\Bigr).
\end{equation}
The resulting force is totally radial; the difference between~\eqref{tidemignard} and~\eqref{Fdiss} is that the second term in the brackets of the first equation, whose direction librates around the radial one, has been averaged in such a way as to obtain the same amount of energy dissipation.

Integrating the equations of motion with the adding of~\eqref{Fdiss}, we obtain values for the variation of the mean motions that are close to~\eqref{dotn}
\[\frac{\dot{n}_1}{n_1}=0.343, \hspace{1.5cm} \frac{\dot{n}_2}{n_2}=-0.306, \hspace{1.5cm}\frac{\dot{n}_3}{n_3}=-1.629.\]

These results have the same magnitude and qualitative behaviour of the ones published in~\cite{laineynature}, which are $0.14$, $-0.43$ and $-1.57$ respectively, although the values are not identical. The possible reason for this disagreement can be a different model of Io's rotation. In our dynamical models we considered Io completely locked in corotation, but other models include physical or geometric librations, such as in~\cite{dirkx}. This is not an aim of this paper to develop a secular theory for the tidal effects in the case of more realistic rotational models, also because the rotation of Io and its rotational parameters are far from being well determined. For what concerns our assumptions, the results are optimal, since in the same condition of corotation given by~\eqref{Fdiss}, we obtain almost the same values from both the semi-analytical model and the numerical one.

Finally, in order to quantify the evolution of the Laplace resonance, we can check how the two-body resonances of the couples Io-Europa and Europa-Ganymede change. Nowadays, the quantity $\nu$, defined in~\eqref{ineq}, is positive and it corresponds to minus the current variation of the longitudes of the pericenters $\varpi_1$ and $\varpi_2$ (almost $-0.7395$ $^\circ/day$). The variation of $\nu$,
\[\dot\nu=\dot n_1 - 2\dot n_2=\dot n_2 - 2\dot n_3,\]
indicates wether the Laplace resonance is going deeper in its state ($\dot\nu<0$), or it is evolving outward ($\dot\nu>0$). From our simulations we obtain $\dot \nu=8.5\times 10^{-8}$ $1/year^2$, which is comparable to $7.4\times 10^{-8}$ of~\cite{laineynature}. In this case the system moves outside the Laplace resonance.

These results depend on the values of the dissipative parameters $k_2^1/Q^1$ and $k_2^0/Q^0$. We chose the last ones published, in order to compare the accelerations we found with the ones of the paper. They were obtained from a fit of $100$ years of astrometric observations; it is possible that with new observations, for example with the extremely precise data of the JUICE space mission as tested in~\cite{dirkxlari}, a new estimation of the parameters can lead to a different evolution of the Laplace resonance.

\section{Conclusion}
In this work we presented a semi-analytical model of the Galilean satellites' dynamics devoted to the description of their resonant and secular motion. From \figurename~\ref{fig:ecc100},~\ref{fig:inc100},~\ref{fig:semiaxe100},~\ref{fig:res100} and \tablename~\ref{tab:residual} we showed that its integration is in good agreement with the filtered series of the JUP310 ephemerides. In particular, as we can see from the values in the table, the standard deviation of the residuals is quite limited, considering a total time of $100$ years. We compared also the main frequencies and secular periods with other works in literature, obtaining similar values. The most satisfying result was to capture with a good precision the frequencies and the amplitudes of the resonant angles' librations.

Moreover, we investigated the tidal dissipation's effects, obtaining almost the same results for the migration of the orbits. As we stated in Section 7, it could be worth to implement different rotational models of Io, in order to align with~\cite{laineynature} or more realistic models. However, this is not straightforward, also for the limited knowledge of the Io's internal structure, and we will consider it for future works.

The main advantage of a semi-analytical model is to have under control all the terms of the dynamics, written in an explicit and simple form in the Hamiltonian. All the frequencies in the output can be explained looking directly at the dependences of the various terms. We can check also the effect of each term and study its contribution in the propagation. In the paper we highlighted the effect of the term $e_3e_4\cos(\varpi_4-\varpi_3)$ to explain the common secular signal of the eccentricities of Ganymede and Callisto, but there are other examples that could be interesting to treat. Therefore, while numerical models are a sort of black box, which provides very precise description of the motion, but do not clarify the various aspects of the dynamics, we implemented a model that can explain the main features of the motion.

Generally, analytical and semi-analytical models have also the advantage to be faster in the computation. Unfortunately, in this case, since at each step we have to recalculate some Laplace coefficients, we lose part of this speed.

Thanks to its good correspondence with the moons' real motion, we can use this model also for longer propagations, in order to investigate the evolution and the origin of the system. Since the dissipation depends on the Io's eccentricity, it is not trivial to determine the orbital evolution looking just at the current values of the orbits' migration.

There are other aspects of the dynamics we can add to the model in order to improve it, for example the precession of the Jupiter's pericenter and node, or the perturbation due to the Galilean satellites' oblateness. However, it is worth noting that out of the resonance it loses its interest and it is necessary a different approach.

This model can be replied for other satellites or planetary systems, paying attention to the problem of the direct and indirect parts of the Laplace coefficients in the case of resonances different from $2:1$. Also the formulation we used for the tidal dissipation can be easily introduced in other contexts, both for analytical and numerical models.

\appendix

\section{Hamiltonian}
\label{completeham}
In this appendix we report the Hamiltonian function that defines the semi-analytical model.\\

(1) Keplerian term
{\tiny
\begin{align*}
\mathcal H_0 &= - \mu_1^2\beta_1^3/(2L_1^2) - \mu_2^2\beta_2^3/(2L_2^2) - \mu_3^2\beta_3^3/(2L_3^2) + n_sL_s
\end{align*}
}
\noindent For the sake of simplicity we use $L_i$ instead of their expressions in terms of the new variables. The last addendum describe the dynamics of the Sun: $n_s$ is its mean motion, which is constant, and $L_s$ is the momentum relative to its longitude.\\

(2) Jupiter's gravitational field
{\tiny
\begin{align*}
 & \mathcal H_{obl} = j_{1;1}/L_1^6 + j_{2;1}/L_2^6 + j_{3;1}/L_3^6 \\
  &  + j_{1;2}(x_1^2+y_1^2-u_1^2-v_1^2) + j_{2;2}(x_2^2+y_2^2-u_2^2-v_2^2) \\
  &  + j_{3;2}(x_3^2+y_3^2-u_3^2-v_3^2) + j_{4;2}(x_4^2+y_4^2-u_4^2-v_4^2) \\
  &  + j_{1;3}/L_1^{10} + j_{2;3}/L_2^{10} + j_{3;3}/L_3^{10} \\
  &  + j_{1;4}(x_1^2+y_1^2-u_1^2-v_1^2) + j_{2;4}(x_2^2+y_2^2-u_2^2-v_2^2) \\
  &  + j_{3;4}(x_3^2+y_3^2-u_3^2-v_3^2) + j_{4;4}(x_4^2+y_4^2-u_4^2-v_4^2)
\end{align*}
}
\noindent In the coefficients $j_{i;l}$, $i$ indicates the Galilean satellite involved and $l$ is the number of the coefficient. \\

(3) Mutual perturbation satellites (up to the second order)
{\tiny
\allowdisplaybreaks
\begin{align*}
 & \mathcal H_{mut} =  a_{1,2;1} \\
  &  + a_{1,2;2}(x_1^2+y_1^2) + a_{1,2;3}(x_2^2+y_2^2) + a_{1,2;4}(x_1x_2+y_1y_2) \\
  &  + a_{1,2;5}(u_1^2+v_1^2) + a_{1,2;6}(u_2^2+v_2^2) + a_{1,2;7}(u_1u_2+v_1v_2) \\
  &  + a_{1,2;8}x_1/\sqrt{L_1} + a_{1,2;9}x_2/\sqrt{L_2} \\
  &  + a_{1,2;10}(x_1^2-y_1^2) + a_{1,2;11}(x_2^2-y_2^2) + a_{1,2;12}(x_1x_2-y_1y_2) \\
  &  + a_{1,2;13}(u_1^2-v_1^2) + a_{1,2;14}(u_2^2-v_2^2) + a_{1,2;15}(u_1u_2-v_1v_2) \\
  &  + a_{2,3;1}(L_2/L_3) \\
  &  + a_{2,3;2}(x_2^2+y_2^2) + a_{2,3;3}(x_3^2+y_3^2) + a_{2,3;4}[(x_2x_3+y_2y_3)cos(\gamma_1)-(x_3y_2-x_2y_3)sin(\gamma_1)] \\
  &  + a_{2,3;5}(u_3^2+v_3^2) + a_{2,3;6}(u_2^2+v_2^2) + a_{2,3;7}[(u_2u_3+v_2v_3)cos(\gamma_1)-(u_3v_2-u_2v_3)sin(\gamma_1)] \\
  &  + a_{2,3;8}[x_2cos(\gamma_1)-y_2sin(\gamma_1)]/\sqrt{L_2} + a_{2,3;9}x_3/\sqrt{L_3} \\
  &  + a_{2,3;10}[(x_2^2-y_2^2)cos(2\gamma_1)-2x_2y_2sin(2\gamma_1)] + a_{2,3;11}(x_3^2-y_3^2) \\
  &  + a_{2,3;12}[(x_2x_3-y_2y_3)cos(\gamma_1)-(x_3y_2+x_2y_3)sin(\gamma_1)] \\
  &  + a_{2,3;13}[(u_2^2-v_2^2)cos(2\gamma_1)-2u_2v_2sin(2\gamma_1)] + a_{2,3;14}(u_3^2-v_3^2) \\
  &  + a_{2,3;15}[(u_2u_3-v_2v_3)cos(\gamma_1)-(u_3v_2+u_2v_3)sin(\gamma_1)] \\
  &  + a_{1,3;1}(L_1/L_3) \\
  &  + a_{1,3;2}(x_1^2+y_1^2) + a_{1,3;3}(x_3^2+y_3^2) + a_{1,3;4}[(x_1x_3+y_1y_3)cos(\gamma_1)-(x_3y_1-x_1y_3)sin(\gamma_1)] \\
  &  + a_{1,3;5}(u_1^2+v_1^2) + a_{1,3;6}(u_3^2+v_3^2) + a_{1,3;7}[(u_1u_3+v_1v_3)cos(\gamma_1)-(u_3v_1-u_1v_3)sin(\gamma_1)] \\
  &  + a_{1,4;1}(L_1/L_4) \\
  &  + a_{1,4;2}(x_1^2+y_1^2) + a_{1,4;3}(x_4^2+y_4^2)  + a_{1,4;4}[(x_1x_4+y_1y_4)cos(\gamma_1+\gamma_2)-(x_4y_1-x_1y_4)sin(\gamma_1+\gamma_2)] \\
  &  + a_{1,4;5}(u_1^2+v_1^2) + a_{1,4;6}(u_4^2+v_4^2) + a_{1,4;7}[(u_1u_4+v_1v_4)cos(\gamma_1+\gamma_2)-(u_4v_1-u_1v_4)sin(\gamma_1+\gamma_2)] \\
  &  + a_{2,4;1}(L_2/L_4) \\
  &  + a_{2,4;2}(x_2^2+y_2^2) + a_{2,4;3}(x_4^2+y_4^2) + a_{2,4;4}[(x_2x_4+y_2y_4)cos(\gamma_1+\gamma_2)-(x_4y_2-x_2y_4)sin(\gamma_1+\gamma_2)] \\
  &  + a_{2,4;5}(u_2^2+v_2^2) + a_{2,4;6}(u_4^2+v_4^2) + a_{2,4;7}[(u_2u_4+v_2v_4)cos(\gamma_1+\gamma_2)-(u_4v_2-u_2v_4)sin(\gamma_1+\gamma_2)] \\
  &  + a_{3,4;1}(L_3/L_4) \\
  &  + a_{3,4;2}(x_3^2+y_3^2) + a_{3,4;3}(x_4^2+y_4^2) + a_{3,4;4}[(x_3x_4+y_3y_4)cos(\gamma_2)-(x_4y_3-x_3y_4)sin(\gamma_2)] \\
  &  + a_{3,4;5}(u_3^2+v_3^2) + a_{3,4;6}(u_4^2+v_4^2) + a_{3,4;7}[(u_3u_4+v_3v_4)cos(\gamma_2)-(u_4v_3-u_3v_4)sin(\gamma_2)]
\end{align*}
}
\noindent In the coefficients $a_{i,k;l}$, $i$ and $k$ indicate the Galilean satellites involved and $l$ is the number of the coefficient. The coefficients with $l=1,8,9$ are not constant, but they depend on the variables $L_i$. \\

(4) Mutual perturbation satellites (third order)
{\tiny
\allowdisplaybreaks
\begin{align*}
& \mathcal H_{mut3}  =  a3_{1,2;1}(x_1^2+y_1^2)x_1 + a3_{1,2;2}(x_2^2+y_2^2)x_1 + a3_{1,2;3}(u_1^2+v_1^2)x_1 + a3_{1,2;4}(u_2^2+v_2^2)x_1 \\
  &      + a3_{1,2;5}(x_1^2+y_1^2)x_2 + a3_{1,2;6}(x_2^2+y_2^2)x_2 + a3_{1,2;7}(u_1^2+v_1^2)x_2 + a3_{1,2;8}(u_2^2+v_2^2)x_2 \\
  &      + b_{1,2;1}(x_1^3-3x_1y_1^2) + b_{1,2;2}[(x_1^2-y_1^2)x_2-2x_1y_1y_2] \\
  &      + b_{1,2;3}[(x_2^2-y_2^2)x_1-2x_2y_2y_1] + b_{1,2;4}(x_2^3-3x_2y_2^2) \\
  &      + b_{1,2;5}[(u_1^2-v_1^2)x_1-2u_1v_1y_1] + b_{1,2;6}[(u_1^2-v_1^2)x_2-2u_1v_1y_2] \\
  &      + b_{1,2;7}[(u_1u_2-v_1v_2)x_1-(v_1u_2+u_1v_2)y_1] + b_{1,2;8}[(u_1u_2-v_1v_2)x_2-(v_1u_2+u_1v_2)y_2] \\
  &      + b_{1,2;9}[(u_2^2-v_2^2)x_1-2u_2v_2y_1] + b_{1,2;10}[(u_2^2-v_2^2)x_2-2u_2v_2y_2] \\
  &      + b_{1,2;11}[(x_1^2-y_1^2)x_2+2x_1y_1y_2] + b_{1,2;12}[(x_2^2-y_2^2)x_1+2x_2y_2y_1] \\
  &      + b_{1,2;13}[(u_1^2-v_1^2)x_1+2u_1v_1y_1] + b_{1,2;14}[(u_1^2-v_1^2)x_2+2u_1v_1y_2] \\
  &      + b_{1,2;15}[(x_1u_2-y_1v_2)u_1+(y_1u_2+x_1v_2)v_1] + b_{1,2;16}[(x_1u_1-y_1v_1)u_2+(y_1u_1+x_1v_1)v_2] \\
  &      + b_{1,2;17}[(u_1u_2-v_1v_2)x_1+(v_1u_2+u_1v_2)y_1] + b_{1,2;18}[(x_2u_2-y_2v_2)u_1+(y_2u_2+x_2v_2)v_1] \\
  &      + b_{1,2;19}[(x_2u_1-y_2v_1)u_2+(y_2u_1+x_2v_1)v_2] \\
  &      + b_{1,2;20}[(u_1u_2-v_1v_2)x_2+(v_1u_2+u_1v_2)y_2] + b_{1,2;21}[(u_2^2-v_2^2)x_1+2u_2v_2y_1] \\
  &      + b_{1,2;22}[(u_2^2-v_2^2)x_2+2u_2v_2y_2] \\
  &      + a3_{2,3;1}(x_1^2+y_1^2)[x_2cos(\gamma_1)-y_2sin(\gamma_1)] + a3_{2,3;2}(x_2^2+y_2^2)[x_2cos(\gamma_1)-y_2sin(\gamma_1)] \\
  &      + a3_{2,3;3}(u_1^2+v_1^2)[x_2cos(\gamma_1)-y_2sin(\gamma_1)] + a3_{2,3;4}(u_2^2+v_2^2)[x_2cos(\gamma_1)-y_2sin(\gamma_1)] \\
  &      + a3_{2,3;5}(x_1^2+y_1^2)x_3 + a3_{2,3;6}(x_2^2+y_2^2)x_3 +a3_{2,3;7}(u_1^2+v_1^2)x_3 + a3_{2,3;8}(u_2^2+v_2^2)x_3 \\
  &      + b_{2,3;1}[(x_2^3-3x_2y_2^2)cos(3\gamma_1)-(3x_2^2y_2-y_2^3)sin(3\gamma_1)] \\
  &      + b_{2,3;2}[((x_2^2-y_2^2)x_3-2x_2y_2y_3)cos(2\gamma_1)-(2x_2y_2x_3+(x_2^2-y_2^2)y_3)sin(2\gamma_1)] \\
  &      + b_{2,3;3}[((x_3^2-y_3^2)x_2-2x_3y_3y_2)cos(\gamma_1)-(2x_3y_3x_2+(x_3^2-y_3^2)y_2)sin(\gamma_1)] + b_{2,3;4}(x_3^3-3x_3y_3^2) \\
  &      + b_{2,3;5}[((u_2^2-v_2^2)x_2-2u_2v_2y_2)cos(3\gamma_1)-(2u_2v_2x_2+(u_2^2-v_2^2)y_2)sin(3\gamma_1)] \\
  &      + b_{2,3;6}[((u_2^2-v_2^2)x_3-2u_2v_2y_3)cos(2\gamma_1)-(2u_2v_2x_3+(u_2^2-v_2^2)y_3)sin(2\gamma_1)] \\
  &      + b_{2,3;7}[((x_2u_2-y_2v_2)u_3-(y_2u_2+x_2v_2)v_3)cos(2\gamma_1)-((y_2u_2+x_2v_2)u_3+(x_2u_2-y_2v_2)v_3)sin(2\gamma_1)] \\
  &      + b_{2,3;8}[((x_3u_2-y_3v_2)u_3-(y_3u_2+x_3v_2)v_3)cos(2\gamma_1)-((y_3u_2+x_3v_2)u_3+(x_3u_2-y_3v_2)v_3)sin(2\gamma_1)] \\
  &      + b_{2,3;9}[((u_3^2-v_3^2)x_2-2u_3v_3y_2)cos(\gamma_1)-(2u_3v_3x_2+(u_3^2-v_3^2)y_2)sin(\gamma_1)] \\
  &      + b_{2,3;10}[(u_3^2-v_3^2)x_3-2u_3v_3y_3] \\
  &      + b_{2,3;11}[((x_2^2-y_2^2)x_3+2x_2y_2y_3)cos(2\gamma_1)-(2x_2y_2x_3-(x_2^2-y_2^2)y_3)sin(2\gamma_1)] \\
  &      + b_{2,3;12}[((x_3^2-y_3^2)x_2+2x_3y_3y_2)cos(\gamma_1)-(2x_3y_3x_2-(x_3^2-y_3^2)y_2)sin(\gamma_1)] \\
  &      + b_{2,3;13}[((u_2^2-v_2^2)x_2+2u_2v_2y_2)cos(\gamma_1)-(2u_2v_2x_2-(u_2^2-v_2^2)y_2)sin(\gamma_1)] \\
  &      + b_{2,3;14}[((u_2^2-v_2^2)x_3+2u_2v_2y_3)cos(2\gamma_1)-(2u_2v_2x_3-(u_2^2-v_2^2)y_3)sin(2\gamma_1)] \\
  &      + b_{2,3;15}[(x_2u_3-y_2v_3)u_2+(y_2u_3+x_2v_3)v_2] \\
  &      + b_{2,3;16}[((x_2u_2-y_2v_2)u_3+(y_2u_2+x_2v_2)v_3)cos(2\gamma_1)-((y_2u_2+x_2v_2)u_3-(x_2u_2-y_2v_2)v_3)sin(2\gamma_1)] \\
  &      + b_{2,3;17}[(u_2u_3-v_2v_3)x_2-(v_2u_3+u_2v_3)y_2] \\
  &      + b_{2,3;18}[((x_3u_3-y_3v_3)u_2+(y_3u_3+x_3v_3)v_2)cos(\gamma_1)+((y_3u_3+x_3v_3)u_2-(x_3u_3-y_3v_3)v_2)sin(\gamma_1)] \\
  &      + b_{2,3;19}[((x_3u_2-y_3v_2)u_3+(y_3u_2+x_3v_2)v_3)cos(\gamma_1)-((y_3u_2+x_3v_2)u_3-(x_3u_2-y_3v_2)v_3)sin(\gamma_1)] \\
  &      + b_{2,3;20}[((u_3u_2-v_3v_2)x_3+(v_3u_2+u_3v_2)y_3)cos(\gamma_1)-((v_3u_2+u_3v_2)x_3-(u_3u_2-v_3v_2)y_3)sin(\gamma_1)] \\
  &      + b_{2,3;21}[((u_3^2-v_3^2)x_2+2u_3v_3y_2)cos(\gamma_1)+(2u_3v_3x_2-(u_3^2-v_3^2)y_2)sin(\gamma_1)] \\
  &      + b_{2,3;22}[(u_3^2-v_3^2)x_3+2u_3v_3y_3] \\
  &      + b_{1,3;1}[(x_1^3-3x_1y_1^2)cos(2\gamma_1)-(3x_1^2y_1-y_1^3)sin(2\gamma_1)] \\
  &      + b_{1,3;2}[((x_1^2-y_1^2)x_3-2x_1y_1y_3)cos(\gamma_1)-(2x_1y_1x_3+(x_1^2-y_1^2)y_3)sin(\gamma_1)] \\
  &      + b_{1,3;3}[(x_3^2-y_3^2)x_1-2x_3y_3y_1] + b_{1,3;4}[(x_3^3-3x_3y_3^2)cos(\gamma_1)+(3x_3y_3^2-y_3^3)sin(\gamma_1)] \\
  &      + b_{1,3;5}[((u_1^2-v_1^2)x_1-2u_1v_1y_1)cos(2\gamma_1)-(2u_1v_1x_1+(u_1^2-v_1^2)y_1)sin(2\gamma_1)] \\
  &      + b_{1,3;6}[((u_1^2-v_1^2)x_3-2u_1v_1y_3)cos(\gamma_1)-(2u_1v_1x_3+(u_1^2-v_1^2)y_3)sin(\gamma_1)] \\
  &      + b_{1,3;7}[((x_1u_1-y_1v_1)u_3-(y_1u_1+x_1v_1)v_3)cos(\gamma_1)-((y_1u_1+x_1v_1)u_3+(x_1u_1-y_1v_1)v_3)sin(\gamma_1)] \\
  &      + b_{1,3;8}[(x_3u_3-y_3v_3)u_1-(y_3u_3+x_3v_3)v_1] + b_{1,3;9}[(u_3^2-v_3^2)x_1-2u_3v_3y_1] \\
  &      + b_{1,3;10}[((u_3^2-v_3^2)x_3-2u_3v_3y_3)cos(\gamma_1)+(2u_3v_3x_3+(u_3^2-v_3^2)y_3)sin(\gamma1)];
\end{align*}
}
\noindent In the coefficients $a3_{i,k;l}$ and $b_{i,k;l}$, $i$ and $k$ indicate the Galilean satellites involved and $l$ is the number of the coefficient.\\

(5) Sun's perturbation
{\tiny
\allowdisplaybreaks
\begin{align*}
& \mathcal H_{sun}  =  s_{1;1}(L_1/L_s) \\
  &  + s_{1;2}(x_1^2+y_1^2) + s_{1;3}[(x_1k_s-y_1h_s)cos(\gamma_1+\gamma_2)-(k_sy_1+x_1h_s)sin(\gamma_1+\gamma_2)] \\
  &  + s_{1;4}(u_1^2+v_1^2) + s_{1;5}[(u_1q_s-v_1p_s)cos(\gamma_1+\gamma_2)-(q_sv_1+u_1p_s)sin(\gamma_1+\gamma_2)] \\
  &  + s_{1;6}[cos(\lambda_{s}+\gamma_1+\gamma_2)x_1-sin(\lambda_{s}+\gamma_1+\gamma_2)y_1] \\
  &  + s_{1;7}[cos(2\lambda_{s}+2\gamma_1+2\gamma_2)(x_1^2-y_1^2)-sin(2\lambda_{s}+2\gamma_1+2\gamma_2)(2x_1y_1)] \\
  &  + s_{1;8}[cos(2\lambda_{s}+\gamma_1+\gamma_2)(k_sx_1+h_sy_1)-sin(2\lambda_{s}+\gamma_1+\gamma_2)(k_sy_1-x_1h_s)] \\
  &  + s_{1;9}[cos(2\lambda_{s}+2\gamma_1+2\gamma_2)(u_1^2-v_1^2)-sin(2\lambda_{s}+2\gamma_1+2\gamma_2)(2u_1v_1)] \\
  &  + s_{1;10}[cos(2\lambda_{s}+\gamma_1+\gamma_2)(q_su_1+p_sv_1)-sin(2\lambda_{s}+\gamma_1+\gamma_2)(q_sv_1-u_1p_s)] \\
  &  + s_{2;1}(L_2/L_s) \\
  &  + s_{2;2}(x_2^2+y_2^2) + s_{2;3}[(x_2k_s-y_2h_s)cos(\gamma_1+\gamma_2)-(k_sy_2+x_2h_s)sin(\gamma_1+\gamma_2)] \\
  &  + s_{2;4}(u_2^2+v_2^2) + s_{2;5}[(u_2q_s-v_2p_s)cos(\gamma_1+\gamma_2)-(q_sv_2+u_2p_s)sin(\gamma_1+\gamma_2)] \\
  &  + s_{2;6}[cos(\lambda_{s}+\gamma_1+\gamma_2)x_2-sin(\lambda_{s}+\gamma_1+\gamma_2)y_2] \\
  &  + s_{2;7}[cos(2\lambda_{s}+2\gamma_1+2\gamma_2)(x_2^2-y_2^2)-sin(2\lambda_{s}+2\gamma_1+2\gamma_2)(2x_2y_2)] \\
  &  + s_{2;8}[cos(2\lambda_{s}+\gamma_1+\gamma_2)(k_sx_2+h_sy_2)-sin(2\lambda_{s}+\gamma_1+\gamma_2)(k_sy_2-x_2h_s)] \\
  &  + s_{2;9}[cos(2\lambda_{s}+2\gamma_1+2\gamma_2)(u_2^2-v_2^2)-sin(2\lambda_{s}+2\gamma_1+2\gamma_2)(2u_2v_2)] \\
  &  + s_{2;10}[cos(2\lambda_{s}+\gamma_1+\gamma_2)(q_su_2+p_sv_2)-sin(2\lambda_{s}+\gamma_1+\gamma_2)(q_sv_2-u_2p_s)] \\
  &  + s_{3;1}(L_3/L_s) \\
  &  + s_{3;2}(x_3^2+y_3^2) + s_{3;3}[(x_3k_s-y_3h_s)cos(\gamma_2)-(k_sy_3+x_3h_s)sin(\gamma_2)] \\
  &  + s_{3;4}(u_3^2+v_3^2) + s_{3;5}[(u_3q_s-v_3p_s)cos(\gamma_2)-(q_sv_3+u_3p_s)sin(\gamma_2)] \\
  &  + s_{3;6}(cos(\lambda_{s}+\gamma_2)x_3-sin(\lambda_{s}+\gamma_2)y_3) \\
  &  + s_{3;7}[cos(2\lambda_{s}+2\gamma_2)(x_3^2-y_3^2)-sin(2\lambda_{s}+2\gamma_2)(2x_3y_3)] \\
  &  + s_{3;8}[cos(2\lambda_{s}+\gamma_2)(k_sx_3+h_sy_3)-sin(2\lambda_{s}+\gamma_2)(k_sy_3-x_3h_s)] \\
  &  + s_{3;9}[cos(2\lambda_{s}+2\gamma_2)(u_3^2-v_3^2)-sin(2\lambda_{s}+2\gamma_2)(2u_3v_3)] \\
  &  + s_{3;10}[cos(2\lambda_{s}+\gamma_2)(q_su_3+p_sv_3)-sin(2\lambda_{s}+\gamma_2)(q_sv_3-u_3p_s)] \\
  &  + s_{4;2}(x_4^2+y_4^2) + s_{4;3}(x_4k_s-y_4h_s) \\
  &  + s_{4;4}(u_4^2+v_4^2) + s_{4;5}(u_4q_s-v_4p_s) \\
  &  + s_{4;6}[cos(\lambda_{s})x_4-sin(\lambda_{s})y_4] \\
  &  + s_{4;7}[cos(2\lambda_{s})(x_4^2-y_4^2)-sin(2\lambda_{s})(2x_4y_4)] \\
  &  + s_{4;8}[cos(2\lambda_{s})(k_sx_4+h_sy_4)-sin(2\lambda_{s})(k_sy_4-x_4h_s)] \\
  &  + s_{4;9}[cos(2\lambda_{s})(u_4^2-v_4^2)-sin(2\lambda_{s})(2u_4v_4)] \\
  &  + s_{4;10}[cos(2\lambda_{s})(q_su_4+p_sv_4)-sin(2\lambda_{s})(q_sv_4-u_4p_s)]
\end{align*}
}
\noindent In the coefficients $s_{i;l}$, $i$ indicates the Galilean satellite involved and $l$ is the number of the coefficient. The other parameters are the mean longitude of the Sun $\lambda_s$ and its equinoctial elements $h_s=e_s\cos(\varpi_s)$, $k_s=e_s\sin(\varpi_s)$, $q_s=s_s\cos(\Omega_s)$ and $p_s=s_s\sin(\Omega_s)$. The coefficients with $l=1$ are not constant, but they depend on the variables $L_i$.   \\

The list of the coefficients is
{\tiny
\allowdisplaybreaks
\begin{align*}
\mu_{ik}&=-Gm_im_k/a_k \\
j_{i;1} &= -(1/2)\mu_i^4\beta_i^7R_J^2J_2  &  j_{i;2}&=-(3/4)\mu_i^2\beta_i^3(R_J/a_i)^2J_2/L_i^3 \\
j_{i;3} &=  (3/8)\mu_i^7\beta_i^{11}R_J^4J_4         & j_{i;4}&=(15/8)\mu_i^2\beta_i^3(R_J/a_i)^4J_4/L_i^3 \\
a_{i,k;1} &= \mu_{ik}c_{(0,0,0,0,0,0)}        & a_{i,k;2}&= \mu_{ik}c^1_{(0,0,0,0,0,0)}/L_i \\
a_{i,k;3} &= \mu_{ik}c^1_{(0,0,0,0,0,0)}/L_k  & a_{i,k;4}&= \mu_{ik}c_{(0,0,-1,1,0,0)}/\sqrt{L_iL_k} \\
a_{i,k;5} &= \mu_{ik}c^2_{(0,0,0,0,0,0)}/(4L_i)  & a_{i,k;6}&= \mu_{ik}c^2_{(0,0,0,0,0,0)}/(4L_k) \\
a_{i,k;7} &= \mu_{ik}c_{(0,0,0,0,-1,1)}/(4\sqrt{L_iL_k})  & a_{i,k;8}&= \mu_{ik}c_{(-1,2,-1,0,0,0)} \\
a_{i,k;9} &= \mu_{ik}c_{(-1,2,0,-1,0,0)}  & a_{i,k;10}&= \mu_{ik}c_{(-2,4,-2,0,0,0)}/L_i \\
a_{i,k;11} &= \mu_{ik}c_{(-2,4,0,-2,0,0)}/L_k  & a_{i,k;12}&= \mu_{ik}c_{(-2,4,-1,-1,0,0)}/\sqrt{L_iL_k} \\
a_{i,k;13} &= \mu_{ik}c_{(-2,4,0,0,-2,0)}/(4L_i)  & a_{i,k;14}&= \mu_{ik}c_{(-2,4,0,0,0,-2)}/(4L_k) \\
a_{i,k;15} &= \mu_{ik}c_{(-2,4,0,0,-1,-1)}/(4\sqrt{L_iL_k}) & & \\
a3_{i,k;1} &= \mu_{ik}c^1_{(-1,2,-1,0,0,0)}/\sqrt{L_i^3}      & a3_{i,k;2} &= \mu_{ik}c^2_{(-1,2,-1,0,0,0)}/(L_k\sqrt{L_i})  \\
a3_{i,k;3} &= \mu_{ik}c^3_{(-1,2,-1,0,0,0)}/(4\sqrt{L_i^3})  & a3_{i,k;4} &= \mu_{ik}c^3_{(-1,2,-1,0,0,0)}/(4L_k\sqrt{L_i})  \\
a3_{i,k;5} &= \mu_{ik}c^1_{(-1,2,0,-1,0,0)}/(L_i\sqrt{L_k})  & a3_{i,k;6} &= \mu_{ik}c^2_{(-1,2,0,-1,0,0)}/\sqrt{L_k^3} \\
a3_{i,k;7} &= \mu_{ik}c^3_{(-1,2,0,-1,0,0)}/(4L_i\sqrt{L_k})  & a3_{i,k;8} &= \mu_{ik}c^3_{(-1,2,0,-1,0,0)}/(4\sqrt{L_k^3}) \\
b_{i,k;1} &=  \mu_{ik}c_{(3-j,j,-3,0,0,0)}/\sqrt{L_i^3}       & b_{i,k;2} &= \mu_{ik}c_{(3-j,j,-2,-1,0,0)}/(L_i\sqrt{L_k}) \\
b_{i,k;3} &= \mu_{ik}c_{(3-j,j,-1,-2,0,0)}/(L_k\sqrt{L_i})  & b_{i,k;4} &= \mu_{ik}c_{(3-j,j,0,-3,0,0)}/\sqrt{L_k^3} \\
b_{i,k;5} &= \mu_{ik}c_{(3-j,j,-1,0,-2,0)}/(4\sqrt{L_i^3})  & b_{i,k;6} &= \mu_{ik}c_{(3-j,j,-1,0,0,-2)}/(4L_i\sqrt{L_k}) \\
b_{i,k;7} &= \mu_{ik}c_{(3-j,j,-1,0,-1,-1)}/(4L_i\sqrt{L_k})  & b_{i,k;8} &= \mu_{ik}c_{(3-j,j,0,-1,-1,-1)}/(4L_k\sqrt{L_i}) \\
b_{i,k;9} &= \mu_{ik}c_{(3-j,j,0,-1,-2,0)}/(4L_i\sqrt{L_k})  & b_{i,k;10} &= \mu_{ik}c_{(3-j,j,0,-1,0,-2)}/(4\sqrt{L_k^3}) \\
b_{i,k;11} &= \mu_{ik}c_{(-1,2,-2,1,0,0)}/(L_i\sqrt{L_k})  & b_{i,k;12} &= \mu_{ik}c_{(-1,2,1,-2,0,0)}(L_k\sqrt{L_i}) \\
b_{i,k;13} &= \mu_{ik}c_{(-1,2,1,0,-2,0)}/(4\sqrt{L_i^3})  & b_{i,k;14} &= \mu_{ik}c_{(-1,2,0,1,-2,0)}/(4L_i\sqrt{L_k}) \\
b_{i,k;15} &= \mu_{ik}c_{(-1,2,-1,0,1,-1)}/(4L_i\sqrt{L_k})  & b_{i,k;16} &= \mu_{ik}c_{(-1,2,-1,0,-1,1)}/(4L_i\sqrt{L_k}) \\
b_{i,k;17} &= \mu_{ik}c_{(-1,2,1,0,-1,-1)}/(4L_i\sqrt{L_k})  & b_{i,k;18} &= \mu_{ik}c_{(-1,2,0,-1,1,-1)}/(4L_k\sqrt{L_i}) \\
b_{i,k;19} &= \mu_{ik}c_{(-1,2,0,-1,-1,1)}/(4L_k\sqrt{L_i})  & b_{i,k;20} &= \mu_{ik}c_{(-1,2,0,1,-1,-1)}/(4L_k\sqrt{L_i}) \\
b_{i,k;21} &= \mu_{ik}c_{(-1,2,1,0,0,-2)}/(4L_k\sqrt{L_i}) & b_{i,k;22} &= \mu_{ik}c_{(-1,2,1,0,0,-2)}/(4\sqrt{L_k^3}) \\
s_{i;1} &= \mu_{is}c^1_{(0,0,0,0,0,0)}/L_i  & s_{i;2} &= \mu_{is}c_{(0,0,-1,1,0,0)}/\sqrt{L_i} \\
s_{i;3} &= \mu_{is}c^2_{(0,0,0,0,0,0)}/(4L_i) & s_{i;4} &= \mu_{is}c_{(0,0,0,0,-1,1)}/(2\sqrt{L_i}) \\
s_{i;5} &= \mu_{is}c_{(0,1,-1,0,0,0)}/\sqrt{L_i} & s_{i;6} &= \mu_{is}c_{(0,2,-2,0,0,0)}/L_i \\
s_{i;7} &= \mu_{is}c_{(0,2,-1,-1,0,0)}/\sqrt{L_i} & s_{i;8} &= \mu_{is}c_{(0,2,0,0,-2,0)}/(4L_i) \\
s_{i;9} &= \mu_{is}c_{(0,2,0,0,-1,-1)}/(2\sqrt{L_i}) & &
\end{align*}
}
\noindent In the case of the couples Io-Europa and Europa-Ganymede, in the coefficients $b_{i,k;l}$, with $l=1,10$, $j$ is equal to $6$, while for Io-Ganymede, $j$ is equal to $4$.
\begin{acknowledgements}
We thank A. Milani for the important discussions about secular theory and M. Fenucci for the careful reading of the manuscript. This research was funded in part by the Italian Space Agency (ASI).
\end{acknowledgements}

\noindent\small{\textbf{Conflict of interest} The author declares that he has no conflict of interest.}



\end{document}